\documentclass[useAMS,usenatbib]{mn2e}

\usepackage{graphicx,times}
\usepackage{colortbl}
\usepackage{soul,color}
\usepackage[normalem]{ulem}

\def\aap{{A\&A}}		
 
\def\apj{{ApJ}}
\def\apjl{{ApJ}}		
		
\def\aj{{AJ}}
\def\apss{{Ap\&SS}} 
\def\mnras{{MNRAS}}
\def\science{{Science}}

\def\newa{{NewA}}

\def\ltorder{\mathrel{\raise.3ex\hbox{$<$}\mkern-14mu
             \lower0.6ex\hbox{$\sim$}}}
\def\r1{$r_1$}

\title[MOCCA Code for Star Cluster Simulations - {}{{II}}. Comparison with $N$-body Simulations]{MOCCA Code for Star Cluster Simulations - {}{{II}}. Comparison with $N$-body Simulations}
\author[M. Giersz,  D.C. Heggie, J.R. Hurley \& A. Hypki]{Mirek Giersz$^{1}$\thanks{E-mail: mig@camk.edu.pl (MG)}, Douglas C. Heggie$^{2}$, Jarrod R. Hurley$^{3}$  and Arkadiusz Hypki$^{1}$\\
$^{1}$Nicolaus Copernicus Astronomical Center, Polish Academy of Sciences, 
ul. Bartycka 18, 00-716 Warsaw, Poland\\
$^{2}$University of Edinburgh, School of Mathematics and Maxwell Institute for Mathematical Sciences, King's Buildings, Edinburgh EH9 3JZ, UK\\
$^{3}$Centre for Astrophysics \& Supercomputing, Swinburne University of Technology, Hawthorn VIC 3122, Australia}

\begin{document}

\date{Accepted \ldots. Received \ldots; in original form \ldots}

\pagerange{\pageref{firstpage}--\pageref{lastpage}} \pubyear{2002}

\maketitle

\label{firstpage}

\begin{abstract} 
We describe a major upgrade of a Monte Carlo code which has previously
been used for many studies of dense star clusters. We outline the
steps needed in order to calibrate the results of the new Monte Carlo
code against $N$-body simulations for large $N$ systems, up to
$N=200000$. The new version of the Monte Carlo code (called MOCCA), in
addition to the features of the old version, incorporates the direct
Fewbody integrator \citep{FCZR2004} for three- and four-body
interactions, and a new treatment of the escape process based on
\citet{FH2000}. Now stars which fulfil the escape criterion are not
removed immediately, but can stay in the system for a certain time
which depends on the excess of the energy of a star above the escape 
energy. They are called potential escapers. With the addition of the
Fewbody integrator the code can follow all interaction channels which are important for the rate of creation of various types of objects observed in star clusters, and ensures that the energy generation by binaries is treated in a manner similar to the $N$-body model.

There are at most three new parameters which have to be adjusted against
$N$-body simulations for large $N$: two (or one, depending on the chosen
approach) connected with the escape process, and one responsible for
the determination of the interaction probabilities. The values adopted
for the free
parameters have at most a weak dependence on $N$. They allow MOCCA to
reproduce $N$-body results with reasonable precision, not only for the
rate of cluster evolution and the cluster mass distribution, but also
for the detailed distributions of mass and binding energy of
binaries. Additionally, the code can follow the rate of formation of
blue stragglers and black hole - black hole binaries. The code
computes interactions between binaries and single stars up to a
maximum separation $r_{pmax}$, and it is found that MOCCA
needs a rather large value of $r_{pmax}$ to get agreement with the $N$-body simulations.

Except for some limitations such as spherical symmetry, a Monte Carlo code such as MOCCA is at present the most advanced code for simulations of real star clusters. It can follow the cluster evolution in detail comparable to
an $N$-body code, but orders of magnitude faster.

\end{abstract}


\section{Introduction}\label{sec:int}

This is the second paper in a new series of papers in which we attempt
to describe the development of MOCCA (MOnte Carlo Cluster simulAtor) 
and its application to the simulations of star cluster evolution.
The first in the series \citep{HyG2012} described in detail recent 
developments of the previous version of the Monte Carlo code (\citet{GHH2008}, and references therein) and the first results of simulations concerning blue
stragglers (BSS) in an evolving star cluster environment. In this paper, we further develop the code and perform a very detailed comparison with $N$-body simulations of large $N$ stellar systems up to $N = 2 \times 10^5$.

 MOCCA \citep{HyG2012} is at present one of the most
advanced numerical codes for stellar dynamical simulations, and is capable of following the evolution of real star clusters in detail comparable to that of $N$-body simulations, but orders of magnitude faster (several hours for $N = 2 \times 10^6$). The dynamical ingredients of the Monte Carlo code are essentially the same as those described in \citet{Gi1998,Gi2001,Gi2006} and \citet{GHH2008}, whose code embodies several features introduced by \citet{St1986}, whose code was in turn based on that originally devised by \citet{He1971}.  Two main features distinguish  MOCCA from the previous version of the Monte Carlo code: (i) it now incorporates dynamical interactions between binary and single
stars and between pairs of binaries based on the Fewbody integrator
developed by \citet{FCZR2004}; (ii) it replaces the treatment of the
escape process in the static tidal field based on \citet{Ba2001} by
one in accordance with the theory proposed by \citet{FH2000}. The
escape process is not instantaneous any more; an object needs time to
find its way around the Lagrangian point $L_1$ to escape. MOCCA incorporates most of the processes which are important during the
evolution of a stellar system, e.g.  relaxation, the main engine of
dynamical cluster evolution; stellar evolution according to
\citet{Hu2000} for the evolution of single stars, supplemented by the
methods of \citet{Hu2002} for the internal evolution of binary stars
and also a simple approach for colliding stars based on the
McScatter interface  by \citet{HPH2006};  escape  in the static
tidal field of the parent galaxy; direct few body integration to
follow interactions between binaries and single stars and other
binaries; binary formation in 3-body interactions; and optional mass-segregated initial cluster configurations according to \citet{BDK2008} and \citet{SKB2008}.

There are several factors which motivate this work. Star clusters are the focus of many intensive observational campaigns \citep[e.g.][and references therein] {bedinetal2001,bedinetal2003,grindlayetal2001,piottoetal2002,
kaliarietal2003,kafkaetal2004,richeretal2004,andersonetal2006,miloneetal2012},
which are now turning to the examination of the parameters of their
populations of different kinds of binaries, BSS and other ``peculiar"
objects. Dynamical models are needed for the design and interpretation
of observational programmes: how is the period distribution and the
spatial distribution of binaries affected by dynamical evolution?
What is the influence of environment and dynamical evolution on the
formation of ``peculiar" objects? Understanding the abundance, spatial
distribution and channels of formation of BSS can only be attempted by
a technique which follows simultaneously both binary dynamics and
internal evolution. While the $N$-body technique may ultimately be the
method of choice for such studies, systems of the size of most
globular clusters are likely to remain beyond reach for some years,
simply because of the number of stars and the size of the binary
population. After all, it is only recently that the ``hardest" open
clusters such as M67 \citep{Hurleyetal2005} and the ``easiest", loosely bound and
distant globular cluster Palomar 14 \citep{Zonoozietal2011} have been
modelled at the necessary level of sophistication. To efficiently
compute detailed models of large star clusters and to investigate the
influence of initial parameters on a cluster's global and local
observational properties we need a technique which is much faster than
the $N$-body code and at the same time can give the same level of
information about every object in the cluster as the $N$-body code
does. MOCCA is such a technique, and is broadly comparable with a Monte Carlo code developed over many   years, largely independently, by the Northwestern group \citep{ChFUR2010}.

One of the drawbacks of  non-direct techniques (including the Monte
Carlo one) compared to the $N$-body model is the necessity of using
free parameters to try to describe the complexity of physical
processes naturally covered in the direct code. The most important
free parameters (from the point of view of MOCCA) are connected with
the relaxation process (the coefficient $\gamma$  in the Coulomb
logarithm), escape  in the static tidal field and, finally, dynamical
interactions between different objects (where the parameter,
$r_{pmax}$, is the maximum pericentre distance between interacting
objects for which few-body interactions are calculated
explicitly). The usual method of determining the free parameters is a
comparison with the results of $N$-body simulations. For the previous
version of the Monte Carlo code the comparison was done only for small
$N$ systems (up to $N = 24000$). The code was successfully used to
simulate the evolution of several real star clusters: M67
\citep{GHH2008}, M4 \citep{HG2008}, NGC6397 \citep[][]{GH2009,HG2009}
and 47Tuc \citep{GH2011}. Despite those successes there were some
doubts connected with the scaling with $N$  of the escape process, the
implementation of which was based on \citet{Ba2001}. To fully validate
MOCCA it has to be tested for larger $N$, and not only for
the global parameters like evolution of the total cluster mass or
Lagrangian radii, but also for the properties and spatial
distributions of binaries and BSS. That kind of comparison will show
how far we can trust the results of Monte Carlo simulations, and which processes cannot be properly described in the framework of MOCCA.

This paper begins in Sec. 2 with a summary of the features which have been added to the Monte Carlo scheme during the construction of the new version of the code, MOCCA. We also show there how we calibrate the free parameters of MOCCA with results of $N$-body simulations.  Next (Sec. 3) we describe the similarities and differences between MOCCA and $N$-body simulations and discuss the possible reasons for them. The final section summarises our conclusions, and discusses some limitations and future developments of MOCCA.

\section{Technique}\label{sec:tech}

 MOCCA \citep{HyG2012} is an updated version of the Monte
Carlo code developed in \citet{Gi1998,Gi2001,Gi2006,GHH2008}. In
addition to the description of the relaxation process, which is
responsible for the dynamical evolution of the system, it includes
synthetic stellar evolution of single and binary stars (BSE code) using
prescriptions described by \citet{Hu2000} and \citet{Hu2002} and
direct integration procedures for small $N$ subsystems based on the
Fewbody code of \citet{FCZR2004}; this is described in more detail below.  One of the more important updates
is a better description of the escape process according to
\citet{FH2000}. Now the escape of an object from the system is no longer
instantaneous, but takes place after some delay. The theory of
\citet{FH2000} incorporates a number of parameters which they had to
determine empirically, and which depend on the system under
consideration; here we shall determine these parameters by comparing
the results with those of $N$-body simulations. 

One kind of interaction for which Fewbody is not used is the formation
of binaries from an encounter of three single stars. 
  The procedure used in MOCCA for binary formation in
  three-body interactions  is the same as the procedure described in
  great detail in \citet{Gi2001}. It relies on the observation that
  the probability that the masses of the three stars involved in the
  interaction are $m_1$, $m_2$ and $m_3$ is proportional to
  $n_1n_2n_3/n^3$ (where $n_1$, $n_2$, $n_3$ and $n$ are the number
  densities of the three interacting stars and the total number
  density, respectively) and the rate of binary formation is
  proportional to $n_1n_2n_3$. These considerations lead to a formula
  for the probability of binary formation which depends only on the
  local total number density instead of the local number densities of
  each mass involved in the interaction (see equation (7) in
  \citet{Gi2001}). This procedure substantially reduces fluctuations
  in the binary formation rate. 

In MOCCA, to decide what is the outcome of an interaction between
  a binary and a field star or another binary, first we need to check if an interaction is due by computing its probability, and if it is then we execute the direct integration procedure for a small $N$ (3- or 4-body) subsystem to find out the outcome of the interaction. The interaction probability depends, among other things, on the maximum
value of the pericentre distance, $r_{pmax}$, beyond which interactions 
are ignored. The larger this distance the larger the
number of interactions, which are weaker on average. Choosing a proper
value of $r_{pmax}$ is crucial for a balance between the efficiency of
the code and its accuracy; e.g. the number of BSS observed in the
system strongly depends on $r_{pmax}$ in some cases (see
Sec. \ref{sec:free}). 

  Physical collisions between stars involved in 3- or 4-body
  interactions are treated according to a standard prescription given
  in \citet{FCZR2004}. Namely, colliding stars are fully mixed and the
  final size of a star is equal to $3(R_1 + R_2)$, where $R_1$ and
  $R_2$ are the radii of the colliding stars. The consequences of such assumptions for the number of BSS are discussed in \citet{HyG2012}.   

  It is worth noting for clarity that all free parameters connected
  with the relaxation process (namely: the deflection angle - $\beta$,
  the overall time step - $\tau$ and the coefficient in the Coulomb logarithm - $\gamma$) were determined in
\citet{GHH2008}, and all free parameters which are intrinsic to the
BSE and Fewbody codes are given the standard values described in \citet{Hu2000,Hu2002} and \citet{FCZR2004}, respectively and also in \citet{HyG2012}. In this paper only the free parameters connected with the escape process and the interaction probability are calibrated.  

\begin{table*}
\begin{minipage}{120mm}
\caption{Initial conditions for $N$-body simulations}
\begin{tabular}{llll}
\hline
   Cluster & N=24000 (M67) & N=100000 (NGC6397) & N=200000\\
\hline
	Number of single stars & $12000$ & 95000 & 195000\\
	Number of binaries & $12000$ & 5000 & 5000\\
	Binary fraction & $0.5$ & $0.05$ & $0.025$\\
	M(0) & $1.869\times10^4M_\odot$ & $5.177\times10^4M_\odot$ & $1.001\times10^5M_\odot$\\
	Initial model & Plummer & Plummer & Plummer\\
	Initial tidal radius & $32.2$pc & $52.4$pc & $35.8$pc\\
	IMF of stars & Kroupa$^a$ & Kroupa$^a$ & Kroupa$^a$\\
	IMF of binaries & Kroupa$^b$ & Kroupa$^b$ & Kroupa$^b$\\
   Mass ratio & Uniform & Uniform & Uniform\\
	Binary eccentricities & thermal$^c$ & thermal & thermal$^c$\\
	Binary semi-major axes & Uniform$^d$ & Uniform$^d$ & Uniform$^d$\\
	SN kick distribution & Gaussian$^e$ &Gaussian$^e$ & flat$^f$\\
	Metallicity & $0.02$ & $0.001$ & $0.001$\\
\hline
\end{tabular}

$^a$ \citet{Kroupaetal1993} with mass range between $0.1$ and $50M_{\odot}$\\
$^b$ \citet{Kroupaetal1991} eq.(1) with mass range between $0.2$ and $100M_{\odot}$\\
$^c$ thermal distribution modified according to \citet{Hurleyetal2005} eq.(1) \\
$^d$ uniformly distributed in the logarithm in the range $2(R_1+R_2)$ to $50$AU\\
$^e$ Gaussian distribution with $\sigma = 190$ km/s \\
$^f$ uniform distribution with kick velocities between $0$ to $100$ km/s  \citep{HS2012, SH2012}
\label{table:nbody}
\end{minipage}
\end{table*} 

\subsection{Delayed escape}\label{sec:escape}

As was pointed out in \citet{FH2000} and \citet{Ba2001} the process of
escape from a cluster in a steady tidal field is extremely
complicated. Some stars which fulfil the energy criterion for escape
(i.e. the condition that the  energy of the star exceed the critical
energy $E_{crit} = -1.5(GM/r_{t})$, where $G$ is the gravitational
constant, $M$ is the total mass and $r_t$ is the tidal radius, see
\citet{Sp1987}) can  still be trapped inside the potential well. Some
of those stars can be scattered back to lower energy before they
escape from the system \citep{King1959}. These two factors cause
the cluster lifetime to scale nonlinearly with relaxation time for
tidally limited clusters \citep{Ba2001}, in contrast with what would
be expected from the standard theory. The efficiency of this effect
decreases as the number of stars increases. To account for the process
described above in the previous version of the Monte Carlo code an
additional free parameter $\alpha$ was introduced \citep{GHH2008}. The
critical energy for escaping stars was approximated by $E_{crit} =
-x_{tid} (GM/r_{t})$, where $x_{tid} = 1.5 - \alpha (ln(\gamma
N)/N)^{1/4}$; here $\alpha$ was approximated by $2.5$ and $\gamma$ is
the coefficient in the Coulomb logarithm, which we take equal to
$0.11$ for the equal mass case and $0.02$ for the unequal mass case
\citep[see][and references therein]{GHH2008}. This prescription was
also tested by \citet{ChFUR2010} in their Monte Carlo simulations of
star clusters. There are three drawbacks of this approach: (i) the
effective tidal radius for Monte Carlo simulations is $r_{t_{eff}} =
r_{t}/x_{tid}$ and it is smaller than $r_{t}$; therefore, for Monte
Carlo simulations, a system was slightly over-truncated (as measured
by the ratio between the tidal radius and the half-mass radius)
compared to $N$-body simulations; (ii) the escape process is
instantaneous: a star whose energy is greater than the critical energy
is promptly removed from the system; (iii) the coefficient $x_{tid}$
is an explicit function of $N$; its $N$ dependence was calibrated only
for low $N$ systems \citep{GHH2008}, and so one can have some doubts about the rate of system evolution for large $N$.

To overcome the drawbacks which are discussed above, we decided to
apply the theory described in \citet{FH2000}. According to this theory
the time-scale for escape is given by
\begin{equation}
t_e = {{2 y_{tid} \sqrt{6} (G M)^{4/3} \omega^{1/3}}\over {\pi (E - E_{crit})^2}},
\label{eq:te}
\end{equation}     
where $\omega$ is the angular velocity of a cluster around a parent galaxy and $E$ is the energy of a star. $y_{tid}$ is a coefficient, which slightly depends on the system structure, and can be approximated by $0.38$. The probability of escape in a time-step $\Delta t$ of a star with energy greater than $E_{crit}$ is given (in theory) by
\begin{equation}
P_a(\Delta t) = 1 - exp(-\Delta t/t_e),
\label{eq:Pa}
\end{equation}
According to Fig. 9 in \citet{FH2000}, however, equation (\ref{eq:Pa})
matches simulation results very poorly not only for the escape
time scale but also for the overall shape of the escape probability
distribution. Indeed strictly equation (\ref{eq:te})  is known only to give an upper limit to the rate of escape, and they found empirically that the true rate of escape is smaller by about a factor of 10. So we can treat $y_{tid}$ in equation (\ref{eq:te}) as a free parameter and adjust it by comparison with $N$-body simulations.

To better represent the empirical shape of the probability
distribution -- which cannot be properly reproduced by equation (\ref{eq:Pa}), even with an
appropriate choice of $y_{tid}$ -- we decided to try also another, empirical
approximation for the probability distribution of escape times
suggested by \citet{FH2000}.    Here the probability of escape in a time-step $\Delta t$ is given by:
\begin{equation}
P_f(\Delta t) = 1 - (1 +b \tilde{t})^{-c},
\label{eq:Pf}
\end{equation}
where $b$ and $c$ are coefficients which depend slightly on the
structure of the system and $\tilde{t} = \omega \Delta t \tilde{E}^2$,
where $\tilde{E} = (E - E_{crit})/|{E_{crit}}|$. In MOCCA
they are  free parameters which will be chosen by  fitting  results from $N$-body simulations. As a first guess, values equal to $3.0$ and $0.8$ can be adopted \citep{FH2000}. 

In MOCCA, escape has to be treated as a Poisson process,
and so it is easy to implement eq.(\ref{eq:Pa}).  But implementation
of eq.(\ref{eq:Pf}) would strictly require that the probability of
escape in the interval $\Delta t$ be computed from $P_f(t+\Delta t)
- P_f(t)$, where $t$ is computed from the time when a star first
becomes a potential escaper, and not from $P_f(\Delta t)$ as stated.
But up to time $t$ the energy of the star changes by relaxation, a
situation which is not envisaged in the numerical results on which
eq.(\ref{eq:Pf}) is based.  Furthermore, for a time step in which
the probability of escape is very small, both expressions
(\ref{eq:Pf}) and (\ref{eq:Pa}) are
proportional to $\Delta t$, with a constant of proportionality which
is proportional to $1/y_{tid}$ and $b$, $c$, respectively.  It is only for the very few time steps in which the probability of escape is large that there is a significant difference between the two formulae (as implemented in MOCCA).

To model escape in MOCCA  according to \citet{FH2000} we now have to
adjust the free parameters $y_{tid}$ or $b$ and $c$ by comparison with
$N$-body simulations. The big advantages of this approach (compared
  with the $N$-dependent approach of the previous version of the
  code) is that the probability of escape does not explicitly depend
on the number of stars in the system and that the escape process
introduced into MOCCA more closely  follows our understanding
of escape  in $N$-body systems.

\subsection{Probability of interactions}\label{sec:prob}

In the hyperbolic 2-body approximation, the total cross section for interaction between a binary and a star, or another binary, with a pericentre distance less than $r_{pmax}$, is given by
\begin{equation}
\sigma = \pi p^2 = \pi r_{pmax}^2 \left (1 + {{2 G m_{123}}\over{r_{pmax} V^2}}\right ),
\label{eq:sigma}
\end{equation}
where $p$ is the impact parameter, $V$ is the relative velocity
between binary and star or other binary, and $m_{123}$ is the total
mass of interacting objects. The second term in the brackets in
equation (\ref{eq:sigma}) is the so called gravitational focusing
term. The larger the value of $r_{pmax}$, the larger the probability
of interaction. In general, the energy released from interactions
should not depend on $r_{pmax}$ for $r_{pmax} \gg a$, where $a$ is the
binary semi-major axis \citep{He1975}. For large enough $r_{pmax}$
there is a balance between small positive and negative binary binding
energy changes, and so the tail of the distribution of $r_{pmax}$  is not
important from the point of view of binary energy generation. Of
course $r_{pmax}$ cannot be too large because in such a case part of
the interactions will be very similar to the ordinary relaxation
process which is already covered by the Monte Carlo engine.  At the
other extreme, for small enough $r_{pmax}$, practically all
interactions will be resonant, hard and connected with strong energy
generation. In this case a substantial number of interactions with
modest energy changes are missing. Of course, a larger value of
$r_{pmax}$ means a larger impact parameter and a larger number of
interactions in each  time interval. Most of these interactions are
very soft from the point of view of energy generation, but may lead to
relatively large changes of binary eccentricity. This means that there
is also a larger probability, either during the interaction  or
shortly after it, for induced  mass exchange between the binary
components.  Together with stellar/binary evolution, this in turn can
lead  to a larger rate of formation of ``peculiar'' objects (e.g. BSS)
or a larger rate of binary merger or disruption.  (That  this is
indeed the case can be seen in Fig.\ref{fig:Nbs-rp-100k}.)  Therefore,
the determination of $r_{pmax}$ is not very important from the point of
view of the total generation of energy by binaries  and the evolution
of cluster global parameters (e.g. total mass or core and half-mass
radii), but it is important from the point of view of the formation of
many kinds of ``peculiar'' objects which are formed in the interplay
between stellar dynamics and stellar evolution. An interesting effect  of $r_{pmax}$ on the spatial distributions of some binary parameters is discussed in Sec.\ref{sec:bss_bh} below.

To roughly assess the range of values of $r_{pmax}$ we will compare
the cross section given in equation (\ref{eq:sigma}) with integrated
differential cross sections over all possible binding energy changes
for resonant interactions according to Heggie's formulae
\citep[][eq. 6-23]{Sp1987}, for flybys and resonant interactions
according to Spitzer's formulae \citep[][eq. 6-27]{Sp1987} and for
hard binary-binary interactions \citep[][eq. 2.7]{Gaoetal1991}.  The
results are
\begin{equation}
{r_{pmax}\over a} = \cases{\displaystyle{{{2 A_H m_1 m_2}\over {7 \sqrt{3 \pi} m_{12} m_3}}} & $for \ Heggie$ \cr
                           \displaystyle{{{5 A_S \pi m_1 m_2}\over {16 \sqrt{3 \pi} m_{12} m_3}}} & $for  \ Spitzer$ \cr
                           \displaystyle{{104\over {42 \pi}}} & $for \ Gao - equal \ mass \ case$, }
\label{eq:rpmax}
\end{equation} 
where $m_1$ and $m_2$ are the masses of the binary components, $m_{12}
= m_1 + m_2$, $m_3$ is the mass of an approaching star, $a$ is the
binary semi-major axis, and $A_S$ and $A_H$ are coefficients defined
in \citet{Sp1987}.  In the limit of strong gravitational focusing and
equal masses, the ratio $r_{pmax}/a$ is equal to $1.1$, $3.4$ and
$0.8$ for the Heggie, Spitzer and Gao cases, respectively. The average
values of $r_{pmax}/a$ in MOCCA simulations (computed for every
interaction according to equation (\ref{eq:rpmax})) are close to the
above values.     We would like to stress that the expressions given
in equation (\ref{eq:rpmax}) are not used in MOCCA, but are   guides to finding $r_{pmax}$. The empirically optimal value of $r_{pmax}$ will  be discussed in Sec.\ref{sec:free}. Note also that, in the case of binary-binary interactions, we always use for $a$ the larger semi-major axis.

\section{$N$-body -- MOCCA Comparison Results}\label{sec:com}

The detailed comparison between results of $N$-body and MOCCA
simulations will proceed in three steps. First, the best values of
$r_{pmax}$, $y_{tid}$, $b$ and $c$ will be chosen by the comparison of
the total mass, the half mass radius, the core radius,  the binary
number and BSS number. Second, the results presented in \citet{Ba2001}
for small $N$ and single mass systems will be checked - the scaling of
the half-mass time with $N$ and the evolution of the number of
potential escapers. Third, the detailed comparison of binary spatial
and energy distributions, binary binding energy, evolution of
Lagrangian radii, average masses in different parts of the system, etc.,
will be done. The detailed comparison between $N$-body and MOCCA
simulation results will help the reader to independently assess from
our conclusions how well MOCCA can follow $N$-body results and how reliable a code it is.

The data for the comparison with MOCCA simulations comes from
simulations done by Jarrod Hurley for M67 \citep{Hurleyetal2005} with
$N=24000$ and  for NGC6397 \citep{Hurleyetal2008} with $N=100000$, and
a model with $N=200000$ (\citet{HS2012}).  The initial conditions for
the $N$-body simulations are summarised in Tab.\ref{table:nbody}.
Note that the initial model, though described as Plummer, is
  actually truncated at the initial tidal radius.

To minimise the statistical fluctuations connected with the generation
of the initial models, the initial positions, velocities, masses and
binary eccentricities and semi major axes for the MOCCA simulations
are taken directly from the $N$-body simulations. They are provided as
input files. The statistical fluctuations observed in the MOCCA
simulations  are therefore connected only with different sequences of interactions and movements of objects in the systems. Positions, velocities and objects chosen for interactions are picked up according to the Monte Carlo technique.

  For Jarrod Hurley's $N$-body simulations the kick velocities due
  to supernova explosions (SN) for black holes (BH) and neutron stars
  (NS) were assumed the same. For models $N=24000$ and $N=100000$ it
  was a Maxwellian distribution with $\sigma = 190$ km/s and for
  $N=200000$ it was a flat distribution between $0 - 100$ km/s. Those
  distributions are far from the widely accepted prescription for kick
  velocities given in \citet{Hobbs2005} and \citet{Fryer2012}, but
  were used by Jarrod Hurley, and so they were also used in MOCCA for its calibration against $N$-body simulations.     

\subsection{Escape algorithm}\label{sec:scaling}

The probability of escape given in equations (\ref{eq:Pa}) and
(\ref{eq:Pf}) depends, among other factors, on the time, $\Delta
t$. According to the Monte Carlo approach, to scale the time step from Monte Carlo units to $N$-body units \citep{HM1986} one needs to use the  equation
\begin{equation}
\Delta t_{NB} = \Delta t_{MC} {N(t)\over ln(\gamma N(t))},
\label{eq:scale}
\end{equation} 
where $N(t)$ is the actual number of objects in the system and
$\gamma$ is the coefficient in the Coulomb logarithm; for multimass
systems this is taken equal to $0.02$ \citep{GHH2008}.   

\begin{figure}
{\includegraphics[height=10.5cm,angle=270,width=8.5cm]{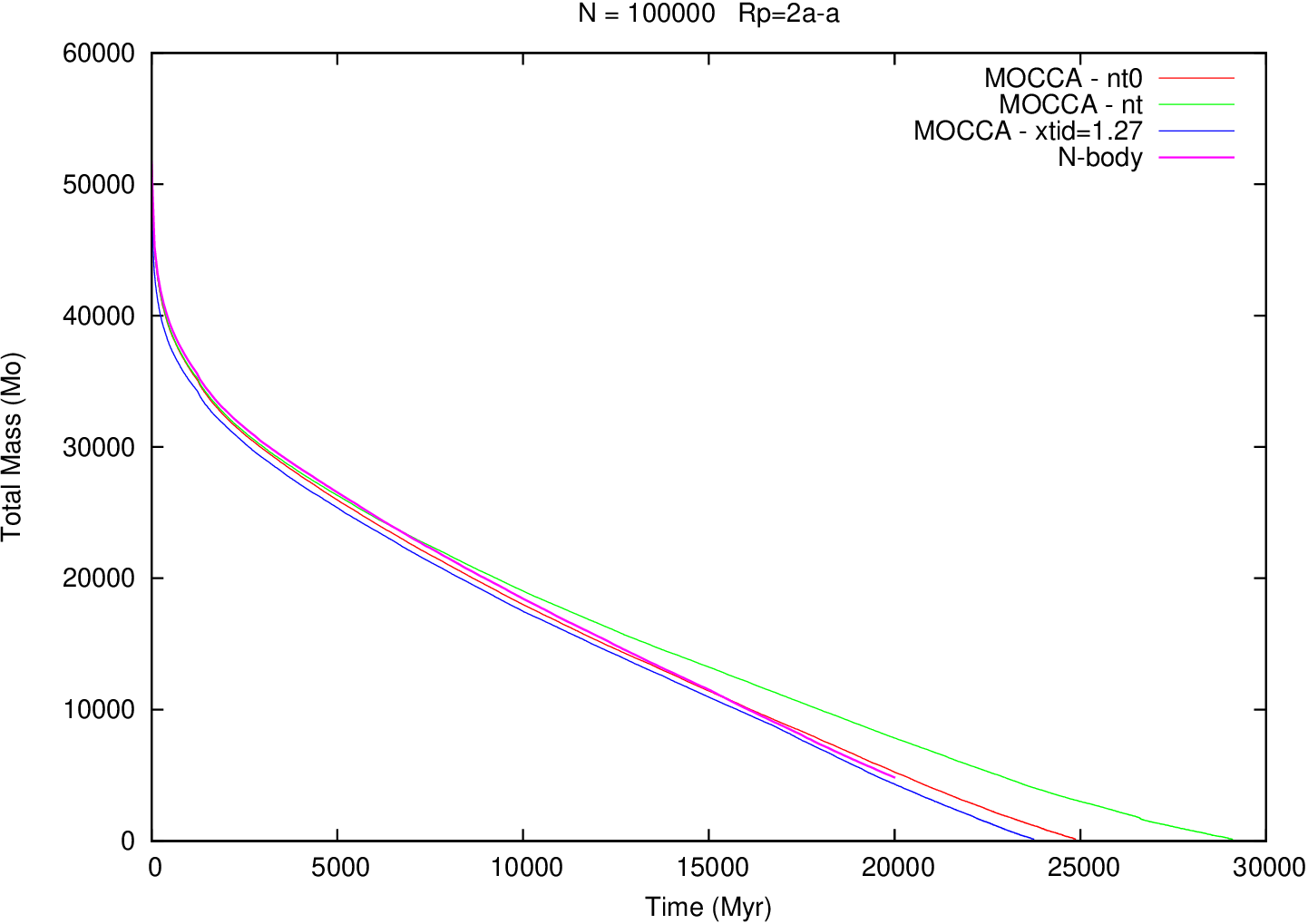}}
    \caption{The evolution of the system total mass as a function of
      time for different MOCCA models and $N$-body
      simulations. MOCCA-nt means scaling between Monte Carlo and
      $N$-body time units according to equation (\ref{eq:scale}) using
      the current value of $N(t)$, MOCCA-nt0 means that, for the
      purposes of determining the escape probability only, scaling
      between Monte Carlo and $N$-body time units uses the initial
      number of objects in the system, $N(0)$, in place of $N(t)$ in
      equation (\ref{eq:scale}), MOCCA-xtid=1.27 means results for
      scaling according to the prescription given in \citet{GHH2008}, and $N$-body means $N$-body results.}
\label{fig:Mnt}
  \end{figure}

As can be seen from Fig.(\ref{fig:Mnt}) the scaling from Monte Carlo
to $N$-body time units (according to equation (\ref{eq:scale})) gives
results which are inconsistent with $N$-body simulations - the
evolution of the total mass is too slow (see the green curve in
Fig.\ref{fig:Mnt}).  This means that the escape rate in the MOCCA
simulations is considerably slower than in $N$-body. The same is true
for  $N$-body simulations with other values of $N$. It has to be
stressed that the two models can not be brought into agreement by
appropriate choice of $y_{tid}$ or $b$ and $c$. Surprisingly,
however, if we compute the escape probability by scaling the time step by a formula like equation (\ref{eq:scale}), but using the initial number of stars $N(0)$ instead of the current value $N(t)$, then it is found that a good match is obtained with $N$-body results, for all values of $N$ that we have checked.

This is a purely empirical finding.   Expressed differently, it
suggests that the value of $y_{tid}$ in equation (\ref{eq:te}) decreases
with time, roughly in proportion to the ratio of the two scalings,
i.e. $N(t)/N(0)$ approximately (if the Coulomb logarithms are
neglected).   The reason for this behaviour is unclear, but the most
plausible explanation is the evolution of the spatial structure of
the system.  According to \citet{FH2000} the parameters in equations
(\ref{eq:Pa}) and (\ref{eq:Pf}) depend on the concentration of the
system:  for more concentrated King models escape is faster.  During
cluster evolution the structure of the system is changing (core --
halo structure is developing), and so one can expect that the
coefficients such as $y_{tid}$ can depend on time.  Preliminary
experiments have shown that, if the value of $y_{tid}$ varies in the
manner suggested by the results of \citet{FH2000} (the ratio of core
to half-mass radius substituting for the King concentration), and if
the correct scaling of equation (\ref{eq:scale}) is restored, then
satisfactory results may be obtained.  But it seems likely that the
escape parameters will also depend on the ratio of the half-mass and
tidal radii, and at present there is no information about this, as
\citet{FH2000} did not study models which underfill the tidal radius.
Further research is needed (and it is currently underway), and
at present we use the alteration of the scaling, even though it
seems inconsistent, as a simple, empirical recipe for incorporating
the time-dependence of the escape parameters. To summarise: in equation (\ref{eq:scale}) we use $N(0)$
instead of $N(t)$ to determine $\Delta t$ for substitution into equation (\ref{eq:Pa}) or (\ref{eq:Pf}), i.e. for the determination of the
escape probability; for all other processes (e.g. stellar evolution, binary interactions or binary formation) to scale time from  Monte Carlo units to physical units, equation (\ref{eq:scale}) was used without alteration.

There are other significant differences between the escape
processes in $N$-body models and in MOCCA, though they are not
apparently time-dependent.  In an $N$-body model the shape of an
equipotential surface  around the tidal radius depends on the value
of the potential energy.   This energy, in the reference frame fixed
to the cluster centre and moving around a host galaxy with angular velocity $\omega$, is given by  
\begin{equation}
E = \cases{\displaystyle{{v^2\over 2} + \phi + {1\over 2}\omega^2(z^2 - 3x^2)} & for $N$-body    \cr
         \displaystyle{{v^2\over 2} + \phi} & for MOCCA,} 
\label{eq:E}
\end{equation}
where $\phi$ is the potential, $v$ is the speed, and $x$ and $z$ are
coordinates with origin at the cluster centre. The last term in the $N$-body expression depends on the centrifugal and tidal forces. In
the case of MOCCA this term is not present, and so the energy
of a star is not exactly comparable in the two methods. In MOCCA
equipotentials  have a spherically symmetrical shape instead of the
approximately triaxial shape in the $N$-body description. This could
lead to some differences in the time to escape between MOCCA and the $N$-body code, which we attempt to overcome by choice of $y_{tid}$ or $b$ and $c$. 

The escape criterion used in the old version of the Monte Carlo code
\citep{GHH2008} gives reasonable agreement with $N$-body results
(despite the fact that it was calibrated only for low $N$), although
the rate of evolution is systematically too fast (see Fig.(\ref{fig:Mnt})). The approximate   agreement confirms that the models of real star clusters computed
with the old Monte Carlo code are relevant. Nevertheless MOCCA, with the new description of the escape process (based on
\citet{FH2000}), gives  results which are more consistent with the
$N$-body results, not only with respect to the evolution of the global
parameters, but also with respect to the detailed properties of binary
distributions. What is also important is that it is $N$-independent and can be safely used for any $N$.

\subsection{Determination of the Free Parameters}\label{sec:free}

To determine the free parameters described above ($y_{tid}$ or $b$ and
$c$) we ran several simulations with  different numbers of stars and
different values for the parameters, and then compared the results with $N$-body simulations. The best values were chosen ``by eye"; we did not attempt to asses how accurate and how unique they are (but see comments about statistical fluctuations below). 
\begin{figure}
{\includegraphics[height=11cm,angle=270,width=9cm]{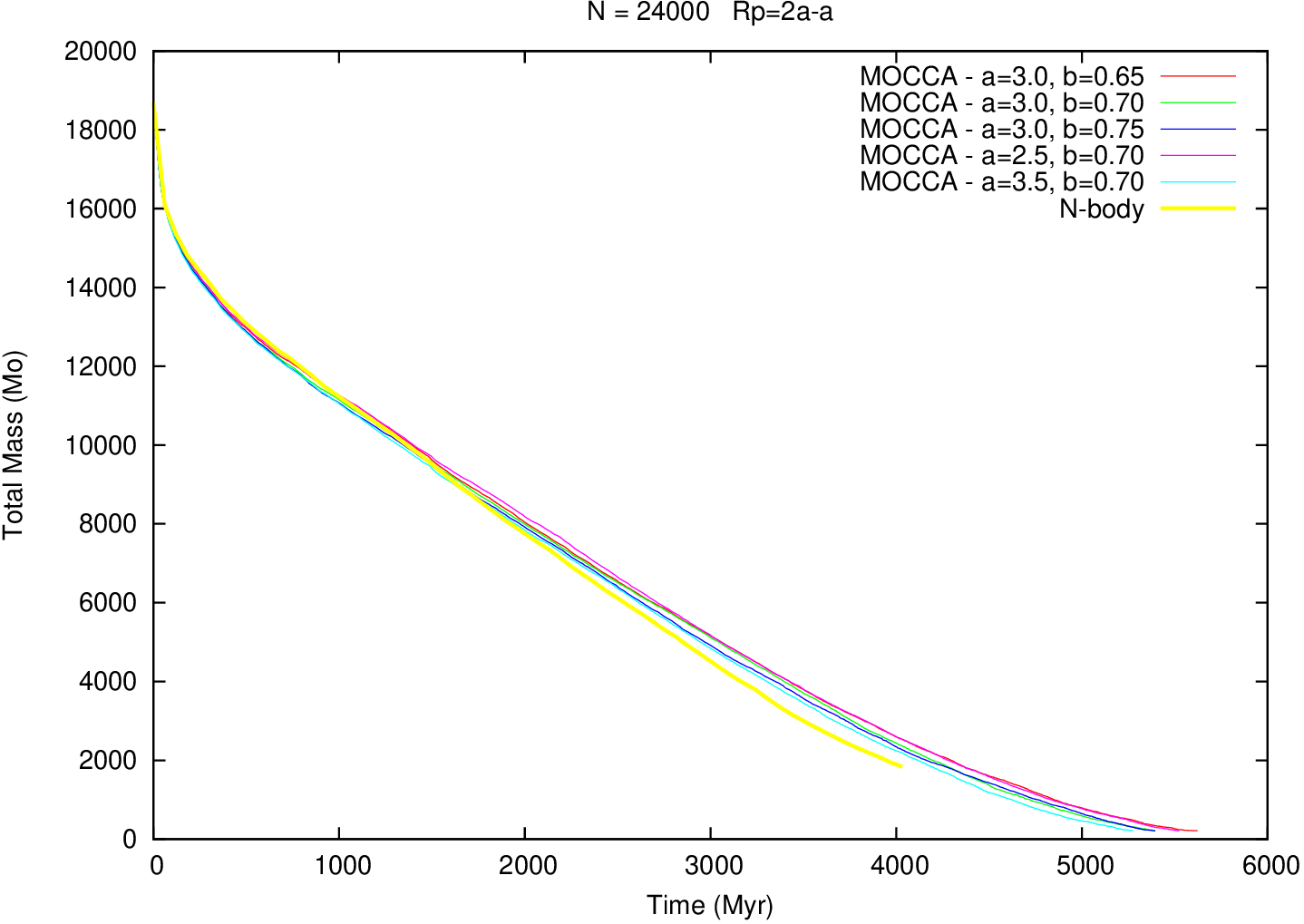}}
    \caption{The evolution of the system total mass as a function of time for MOCCA models with different coefficients $b$ and $c$ and $N$-body simulations for $N=24000$.}
\label{fig:Mab-m67}
  \end{figure}

\begin{figure}
{\includegraphics[height=11cm,angle=270,width=9cm]{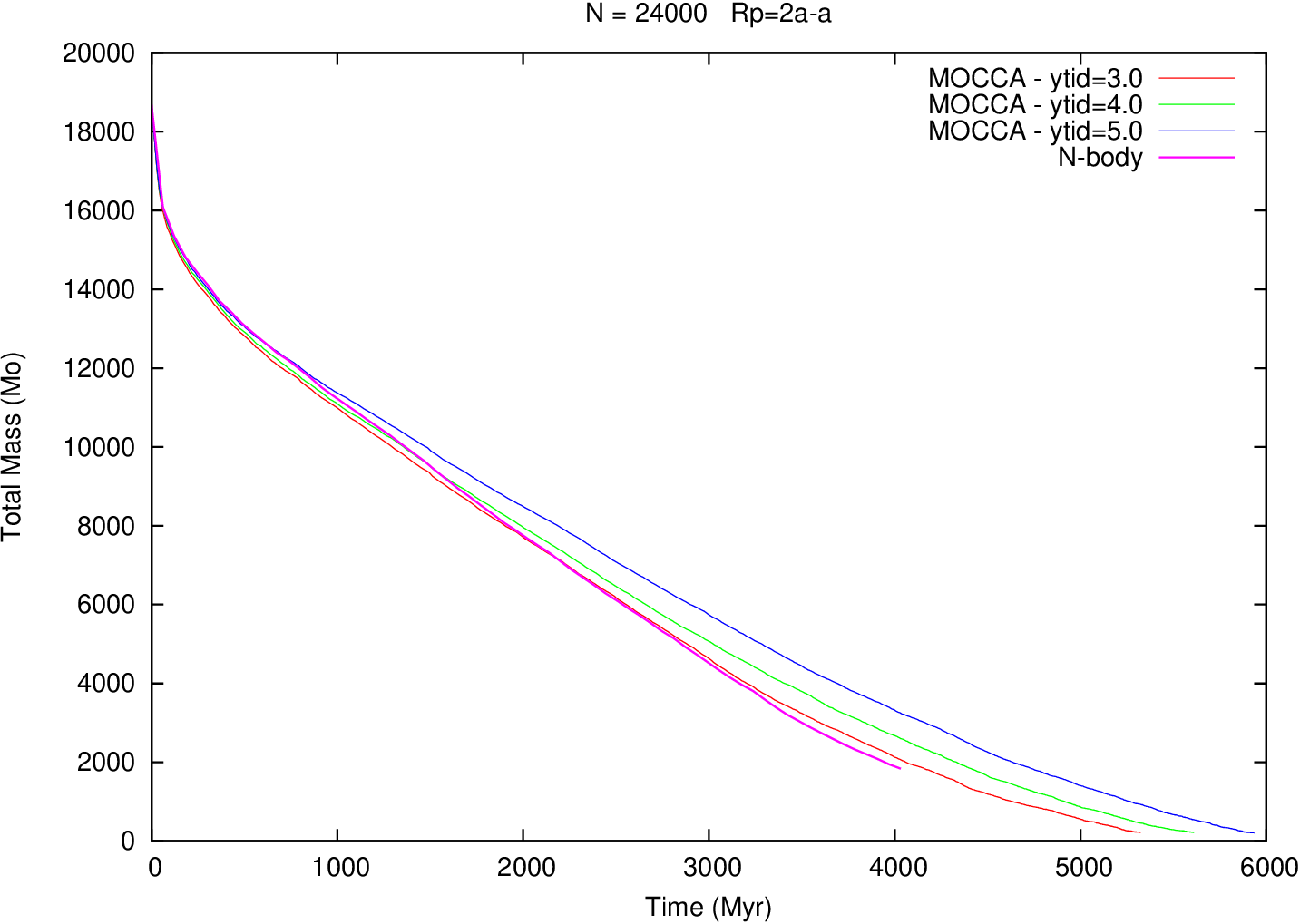}}
    \caption{The evolution of the system total mass as a function of time for MOCCA models with different coefficient $y_{tid}$ and $N$-body simulations for $N=24000$.}
\label{fig:Mytid-m67}
  \end{figure}

\begin{figure}
{\includegraphics[height=11cm,angle=270,width=9cm]{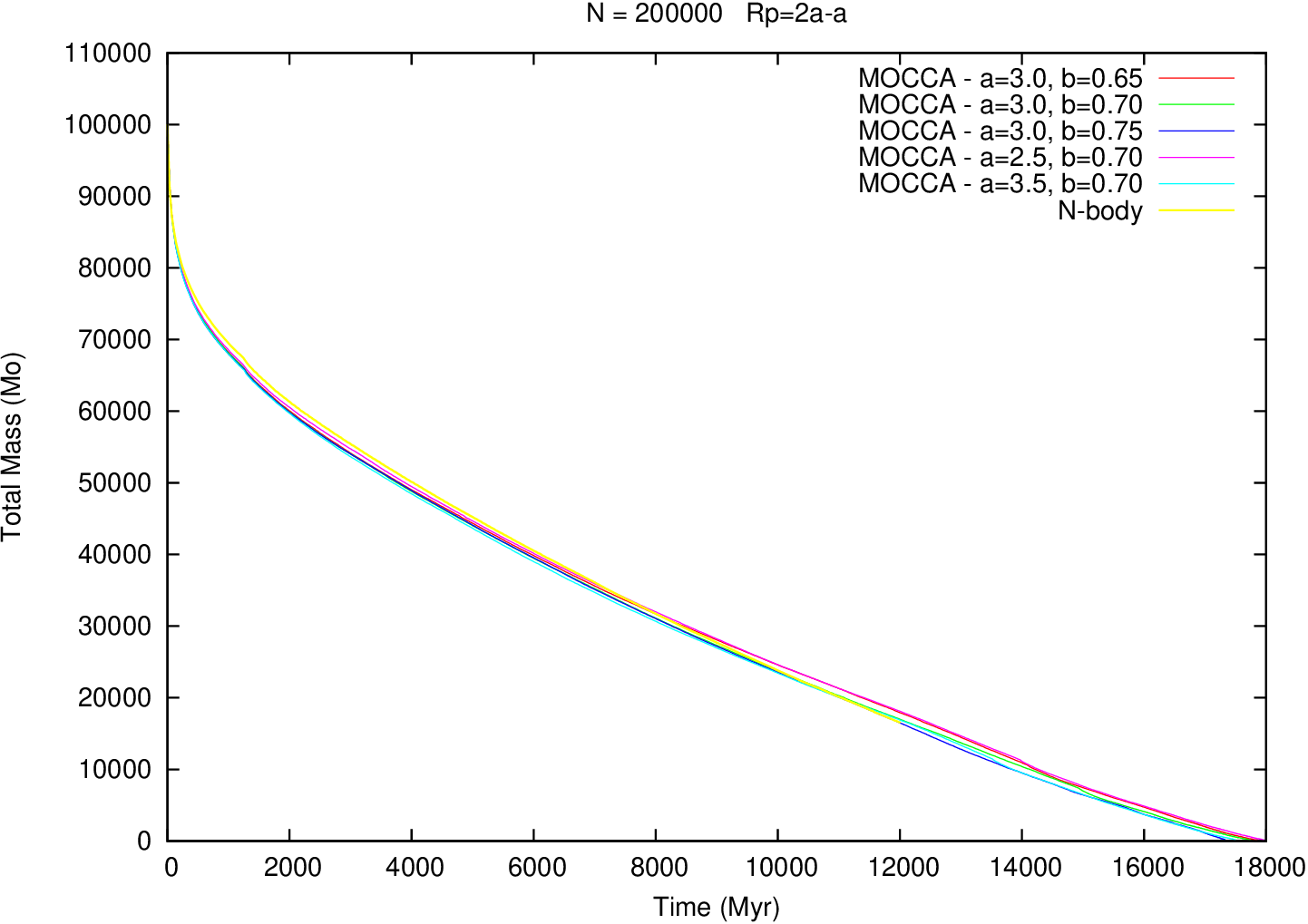}}
    \caption{The evolution of the system total mass as a function of time for MOCCA models with different coefficients $b$ and $c$ and $N$-body simulations for $N=200000$.}
\label{fig:Mab-200k}
  \end{figure}  
The results are presented in Figs.\ref{fig:Mab-m67},
\ref{fig:Mytid-m67} and \ref{fig:Mab-200k}. As one can see, the
dependence on $y_{tid}$ appears to be stronger than on $b$ and
$c$, but the range of values which has been sampled is relatively
  larger for $y_{tid}$  than for $b,c$. The ``best" values are
$y_{tid} = 4.0$ and $b = 3.0$, $c = 0.7$. The values for $b$ and $c$
are close to the values given in \citet{FH2000} in Table 1 there,
which are based on a numerical fit. $y_{tid}$ is about ten times
larger than given in \citet{FH2000} (equation (9) there, which is a
  theoretical estimate), but taking into account their finding that the time scale
given in equation (\ref{eq:Pa}) is too short by comparison with
  numerical data (by about 1 dex), the value $4.0$ corresponds very
closely to the empirical results which \citet{FH2000} found.  Only for
the simulations with $N=24000$ are these values not quite
satisfactory:  the evolution rate is slightly too slow in MOCCA, for
times larger than about 2 Gyr, and it seems that $y_{tid}=3.0$ is a
better choice than $y_{tid}=4.0$. For other $N$ ($100000$ and
$200000$) the MOCCA simulations follow the $N$-body results very well,
and the choice of $y_{tid}=4.0$ gives better agreement.

\begin{figure}
{\includegraphics[height=11cm,angle=270,width=9cm]{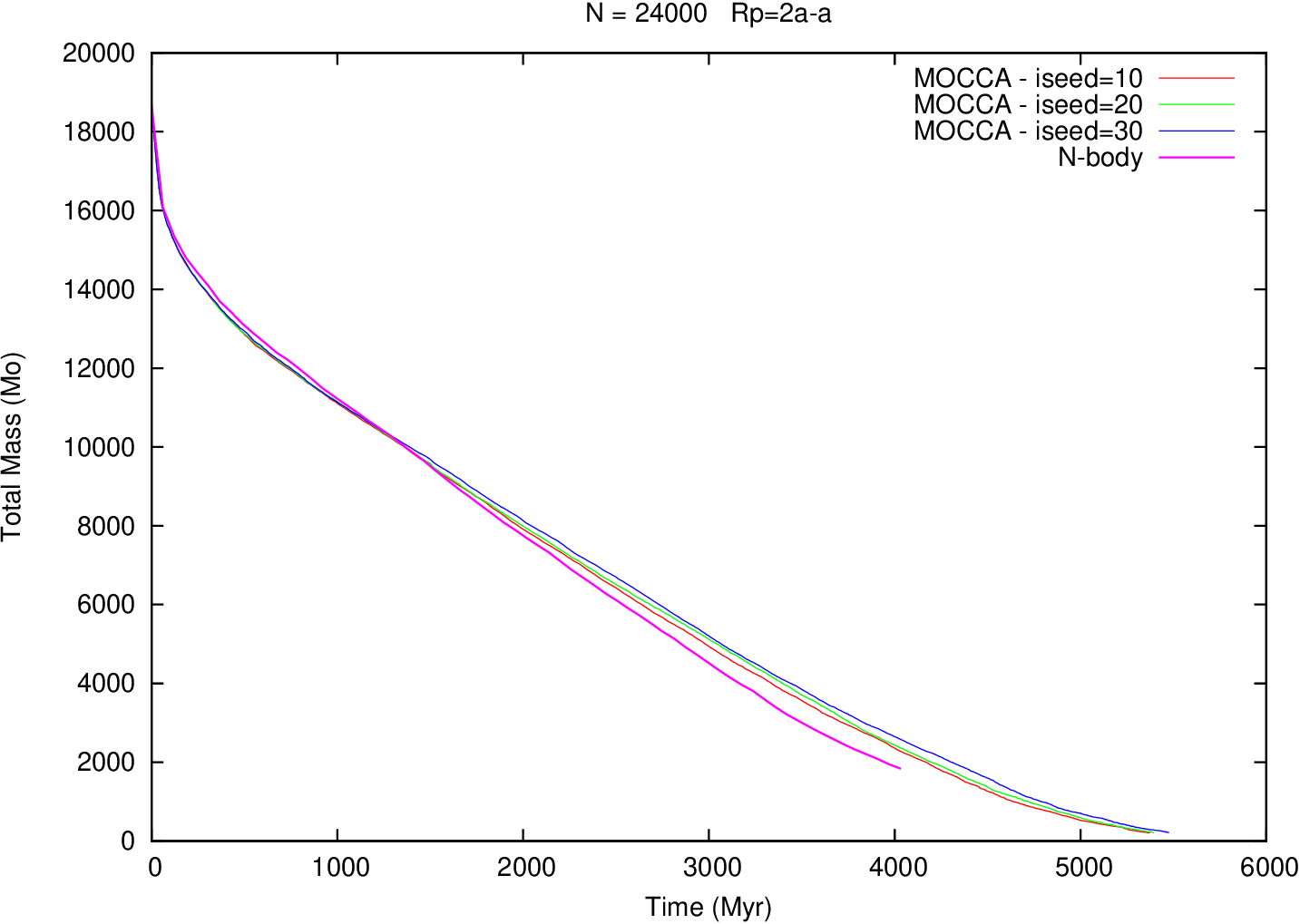}}
    \caption{The evolution of the system total mass as a function of time for MOCCA models with $b=3.0$ and $c=0.7$ and different random sequences (iseed=10, iseed=20, iseed=30) and $N$-body simulations for $N=24000$.}
\label{fig:Mise-m67}
  \end{figure}

To assess the influence of statistical fluctuations on the results
obtained, three simulations with exactly the same initial conditions
but with different sequences of random numbers were run. The results
are given in Fig. \ref{fig:Mise-m67} for $N=24000$. The fluctuations
for this model are largest, but clearly smaller than the difference
connected with different values of the free parameters. For larger $N$
the fluctuations are practically negligible. For other global
quantities the fluctuations are similar to those  for the total
mass. They are small, and the results presented in the paper  for a single simulation are representative.    

\begin{figure}
{\includegraphics[height=11cm,angle=270,width=9cm]{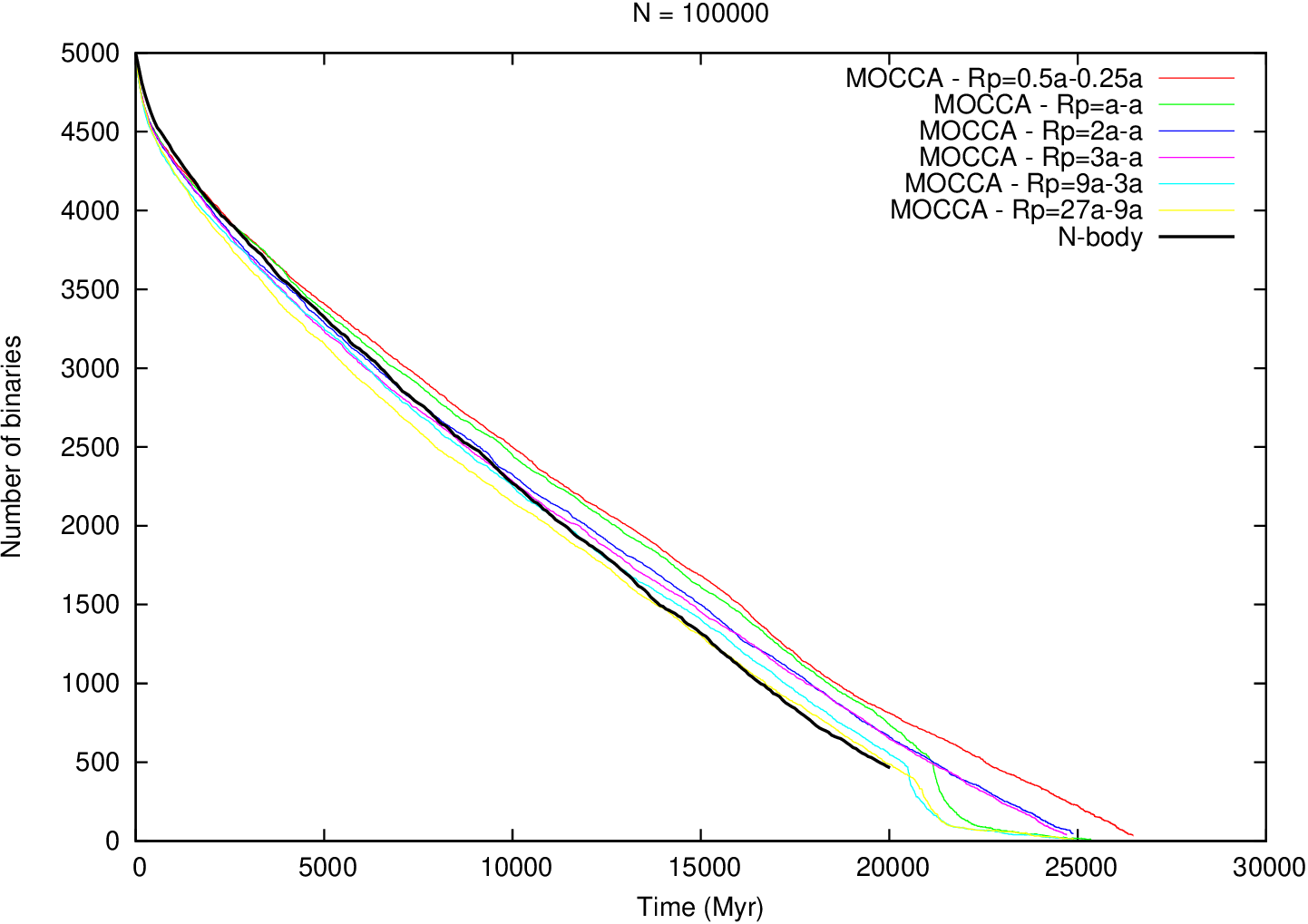}}
    \caption{The evolution of the number of binaries in the system as
a function of time for MOCCA models with $b=3.0$ and $c=0.7$ and different $r_{pmax}$ and $N$-body simulations for $N=100000$.  The notation
Rp=27a-9a, for example, means that $r_{pmax}=27a$ for three-body interactions and $r_{pmax}=9a$ for four-body interactions.}
\label{fig:Nb-rp-100k}
  \end{figure}

\begin{figure}
{\includegraphics[height=11cm,angle=270,width=9cm]{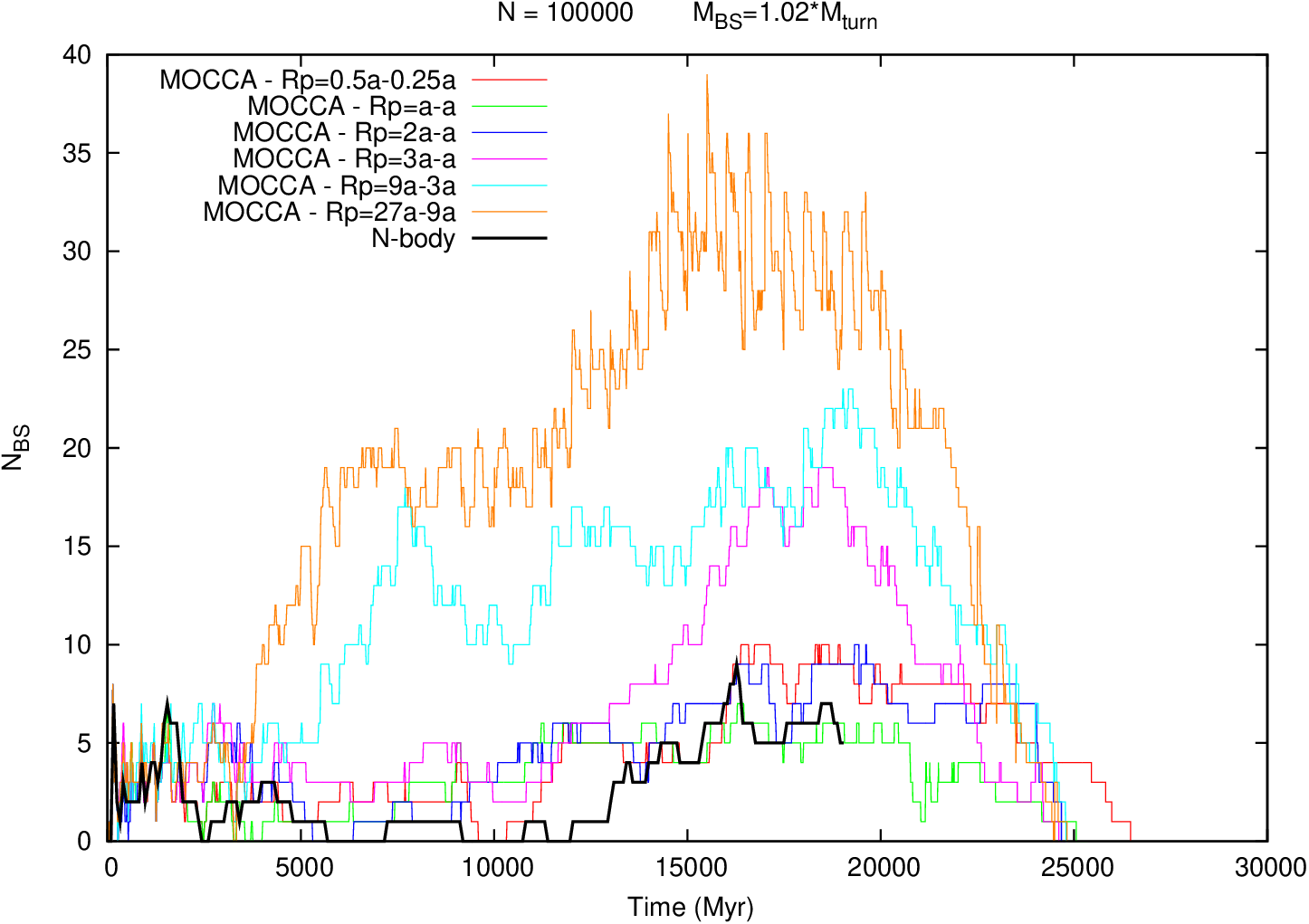}}
    \caption{The evolution of the number of BSS in the system as a
function of time for MOCCA models with $b=3.0$ and $c=0.7$ and different $r_{pmax}$ and $N$-body simulations for $N=100000$.  The notation  Rp=27a-9a,
for example, means that $r_{pmax}=27a$ for three-body interactions and $r_{pmax}=9a$ for four-body interactions.}
\label{fig:Nbs-rp-100k}
  \end{figure}

Having determined $y_{tid}$, $b$ and $c$ we now turn to finding the best
value for $r_{pmax}$. As  was argued above, the value of $r_{pmax}$
will have a big impact on the number and distribution of binaries and
BSS in the system. In MOCCA, BSS are defined exactly as  in
Hurley's $N$-body simulations: a main sequence star is identified as a
BSS when its mass is greater than $1.02 M_{to}$, where $M_{to}$ is the
turn-off mass. As can be seen from Figs. \ref{fig:Nb-rp-100k} and
\ref{fig:Nbs-rp-100k} the requirements set by the $N$-body simulation
for the numbers of BSS and binaries are rather contradictory from the
point of view of comparison with the MOCCA results.  

To get the best agreement for the evolution of the total number of
binaries, MOCCA needs a value of  $r_{pmax}$ which is as large as
possible: $r_{pmax}=27a$ seems to be a good choice. On the other hand
to match the $N$-body results for the number of BSS, MOCCA prefers
modest values of $r_{pmax}$,  equal to about $a$. That conclusion is
also true for $=24000$ and $N=200000$. It seems that a reasonable
compromise between the evolution of the total number of binaries and
BSS is given for the value of $r_{pmax}$ suggested by the theoretical
considerations given above in Sec.\ref{sec:prob}, namely for
three-body interactions $r_{pmax}=2a$ and for four-body interactions
$r_{pmax}=a$. One might suppose that these contradictory requirements
set by the numbers of BSS and binaries could be explained by the
different definitions of binary in $N$-body and MOCCA simulations. In
the $N$-body results presented here, a binary is identified whenever
it is a regularised binary (i.e. a so-called KS binary), and is
therefore rather hard. In MOCCA, however, we follow all binaries,
except extremely soft
ones, which are artificially disrupted in binary-single dynamical
  interactions according to a prescription derived from
  \citet[eq. 4.12]{He1975} (see also the discussion in Sec. 2.4 in
  \citet{HyG2012}). Therefore in MOCCA simulations we should  generally
expect larger numbers of binaries than in $N$-body simulations. To
quantify this we checked the number of non-KS binaries in the $N$-body
simulations. This number was rather small (at most about 30) and in
fact could explain only part of the observed differences.   

It is worth  noting that, as one can expect, for MOCCA the
evolution of the Lagrangian radii and the total mass do not depend on
the value of $r_{pmax}$ - the total energy generation in three- and
four-body interactions does not depend on $r_{pmax}$ provided that
$r_{pmax}$ is not too small or too large. The situation is different
when the Fewbody integrations are switched off in MOCCA and only cross
sections are used for energy generation in binary interactions
(MOCCA-NoFB). Then there is a very strong dependence on $r_{pmax}$ for
the  evolution of Lagrangian radii. Larger $r_{pmax}$ means a larger
probability for interactions. Each interaction generates on average
the same amount of energy (according to the adopted cross section),
and so for larger $r_{pmax}$ more energy is generated by binaries, and
the system expands faster than in the case of MOCCA simulations (with
the Fewbody integrator).
It seems that the best values of $r_{pmax}$ for MOCCA-NoFB are $0.5a$ and $0.25a$,  for three- and four-body interactions, respectively.     

In the remainder of the paper the best values of the free parameters
given in this section, i.e. $b=3.0$, $c=0.7$, $r_{pmax}=2a$ and $a$
for three- and four-body interactions, respectively, will be used for
comparison of different system and binary properties obtained in the MOCCA and $N$-body simulations. The $y_{tid}$ parameter will not be used in the rest of the paper to determine the escape probability.

We would like to stress that all free parameters determined in
this paper were obtained by comparisons with $N$-body simulations
with $N$ from $24000$ up to $200000$. Therefore strictly speaking
they are valid only in this range of $N$. However we give independent comparisons extending to smaller $N$ in Sec.\ref{sec:half}, without changing the values of the free parameters, which suggests that any $N$-dependence of the free
parameters is weak. Therefore we argue that there is no reason to
expect any sudden changes of the parameters for larger $N$. Of course this conclusion will be checked if $N$-body simulations with $N$ of the order of $5 \times 10^5 - 10^6$  become available.

\subsection{Half mass time and potential escapers}\label{sec:half}

Heaving chosen the free parameters $b$ and $c$, which determine the rate of escape from a tidally limited cluster, we can now compare with the $N$-body results of \citet{Ba2001}; in particular, the $N$-dependence of the half-mass time (i.e. the time when the system contains half of its initial mass), and the
evolution of the number of potential escapers  for different $N$. The free parameters determining the time-scale for escape were calibrated in MOCCA against $N$-body simulations of multi-mass systems described by the Plummer model, with stellar evolution and a substantial number of primordial binaries. If the results of the MOCCA simulations follow the results given in \citet{Ba2001}, which were equal-mass systems with no stellar evolution and no primordial binaries, starting with a $W_0 =3$ King model, we will be more assured that the set of free parameters determined in  Sec. \ref{sec:free} give an adequate description of the escape process in a variety of physical models of star clusters. It was for this reason that we decided to use the results of \citet{Ba2001} instead of
\citet{BM2003}, whose cluster models are more similar to the models used for the calibration of MOCCA. Additionally, \citet{Ba2001} provides information about potential escapers, and so we were able to check the intrinsic mechanism of the escape process.   

\begin{figure}
{\includegraphics[height=11cm,angle=270,width=9cm]{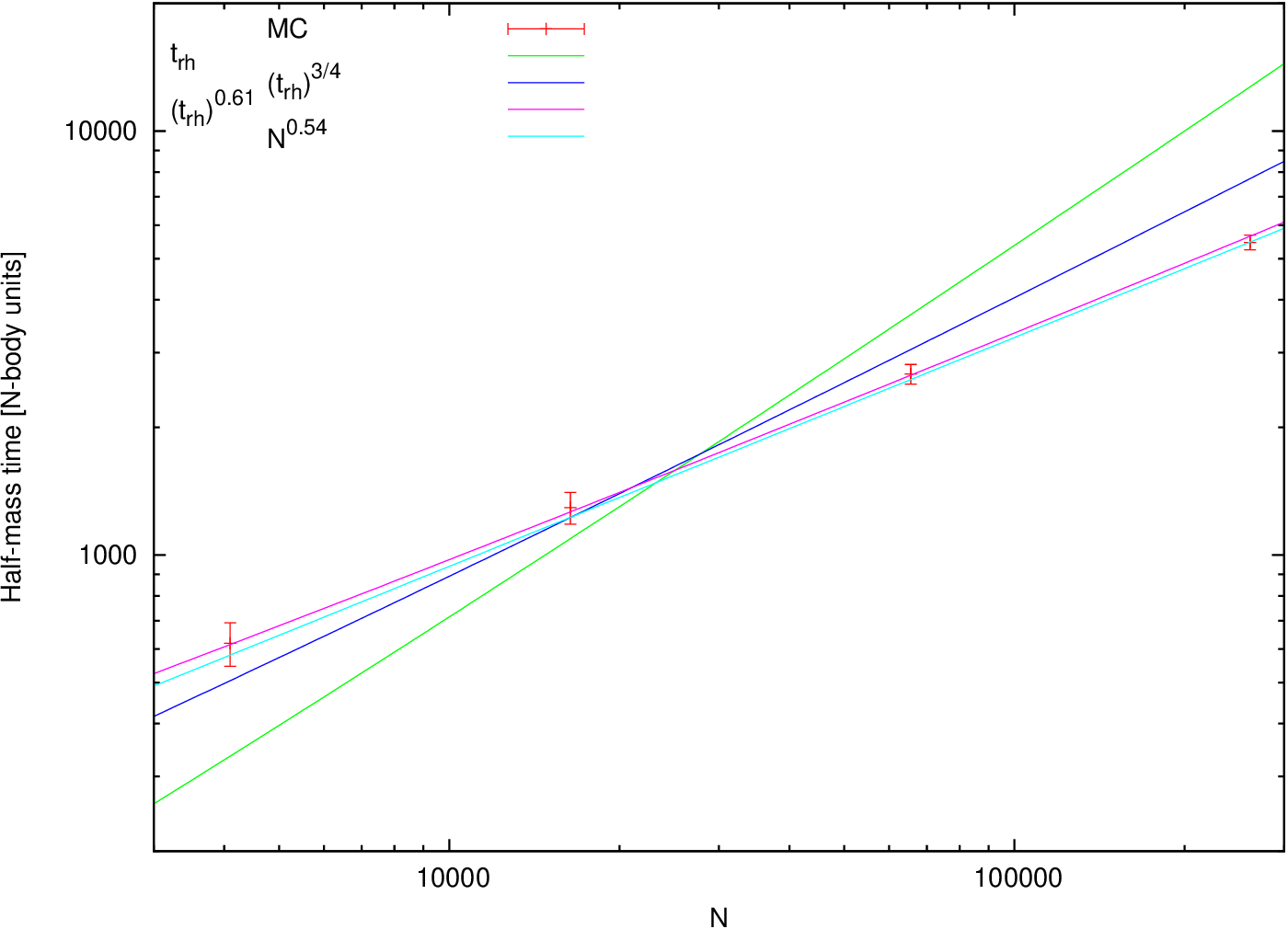}}
    \caption{The dependence of the half mass time on the initial total
      number of stars $N$. Green line: the standard linear scaling
      with the half mass relaxation time, $t_{rh}$; blue line: scaling
      predicted by \citet{Ba2001}, i.e. $t_{rh}^{3/4}$; purple line:
      the scaling obtained in this paper, $t_{rh}^{0.61}$; pale blue line:
      the scaling with $N$ in this paper, $N^{0.54}$. Points with error bars ($3\sigma$) are the results of MOCCA simulations.} 
\label{fig:half}
\end{figure}

According to the model presented by \citet{Ba2001} the half-mass time
should scale as the relaxation time to the power $3/4$ instead of the
linear scaling with the relaxation time predicted by the standard
theory.  Fig.\ref{fig:half}  shows the half-mass time as a function of
$N$ for MOCCA simulations from $N=4K$ to $N=256K$. The figure also shows
different scaling laws fitted to the simulation data. As we can see, the theoretical scaling with $t_{rh}^{3/4}$ increases a little more quickly with $N$ than the MOCCA results.  Baumgardt found a very similar result up to the largest $N$ which he considered (16k), but in the MOCCA results this behaviour extends up to 256k (the largest $N$ considered in this paper). Purely empirically, the simplest accurate fit to our data is a power law: the whole range of $N$ can
be fitted reasonably well with a scaling proportional to $t_{rh}^{0.61}$ or simply $N^{0.54}$.  Our most significant contribution here is to note that the same power law in $N$ (to two significant figures in the power) gives the best power-law fit to the $N$-body results of \citet[Table 1]{Ba2001} for the case of a cluster in a full tidal field. This result shows that, in terms of mass loss, the behaviour of MOCCA agrees remarkably well with $N$-body data.  The fact that the best fitting power law is somewhat different from theoretical expectation
is an issue for theory, and not for these numerical methods. It is worth emphasising that, according to the results presented by \citet{BM2003} the power-law index for the half-mass time depends on the cluster concentration. The larger the concentration the larger the value of the power-law index. It depends even more strongly on whether the cluster is Roche-lobe filling \citep{TF2005}.  Therefore our results depend on the initial conditions we adopted, following \citet{Ba2001}.

\begin{figure}
{\includegraphics[height=11cm,angle=270,width=9cm]{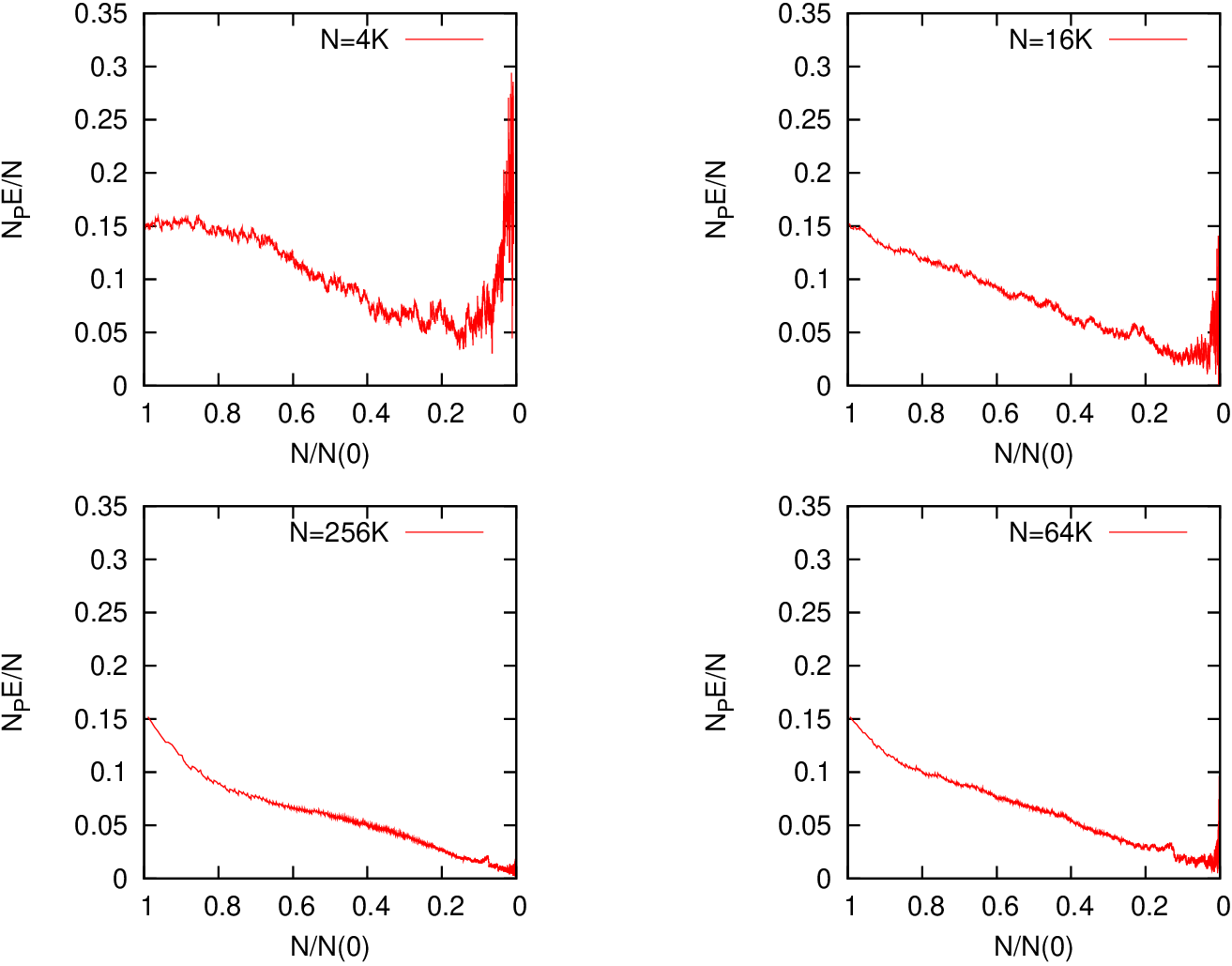}}
    \caption{Potential escaper fraction as a function of stars bound to the cluster for $N=4K$, $N=16K$, $N=64K$ and $N=256K$, clockwise.}
\label{fig:poten}
\end{figure}

Fig.\ref{fig:poten} shows the evolution of the potential escaper
fraction with time for different initial numbers of stars. The  setup
of the initial model is responsible for the $15$ per cent population
of potential escapers at the beginning. The cluster starts with
primordial escapers because the escape energy, $E_{crit}$, is lower
than the edge potential of the initial King model. The different
behaviour (for different $N$) of the number of potential escapers  at
the beginning is in agreement with the results presented by
\citet{Ba2001} (his Fig.11). The initial increase of the number of
potential escapers probably indicates the phase when the cluster
evolves towards equilibrium after removal of a substantial amount of mass
\citep{Ba2001} in a short time. This increase is largest for small
$N$ systems, but for the largest $N$ it is barely visible. The number
of potential escapers decreases with time until  core collapse, when
it starts to rise again. The comparison of Fig.\ref{fig:poten} for
$N=4k$ and $N=16k$ with Fig. 11 in \citet{Ba2001} suggests that  core
collapse is delayed in MOCCA.  Detailed inspection along with Fig. 8
in \citet{Ba2001} shows that the delay is rather modest, not larger
than $5-10$ per cent. The reasons for this can be connected with the
facts that the value of $r_{pmax}=a$ chosen for the simulations was too
small (as is suggested by results presented in Sec. \ref{sec:free}), and that the probability for binary formation in three body interactions was also slightly too small in MOCCA. Indeed, larger $r_{pmax}=27a$ and larger formation probability both bring the  MOCCA results into much better agreement with the $N$-body results, though core collapse in MOCCA is  still slightly delayed.  This suggests that there are other factors which could be responsible for this
disagreement, e.g. the coefficient $\gamma$ in the Coulomb logarithm,
the rate (in both $N$-body codes and in MOCCA) at which 
clusters with small $N$ can regain equilibrium after the initial substantial mass
loss, or the well known fact that the Monte Carlo method is not expected
  to be valid
for  systems with very small $N$, when the crossing time approaches the
half-mass relaxation time. The observed disagreement does not
influence the results for the half mass time and the evolution of the
number of potential escapers before the collapse time. We can conclude
that MOCCA reproduces reasonably well the results presented by \citet{Ba2001} for small $N$-body simulations.

\subsection{Results of comparison}\label{sec:results}

The comparison between MOCCA and $N$-body results will proceed in
three steps. First, the evolution of the global parameters (total
mass, Lagrangian radii, core radius) will be compared. Second, the
evolution of properties of the binaries  (number, energy, mass and number distributions) will be checked. Third,   
properties of  ``peculiar'' objects like BSS and black holes will
be compared. Most of the figures presented below will also display the
results of MOCCA simulations with the Fewbody integrator  switched
off and  interaction cross sections switched on (MOCCA-NoFB). This
will help the reader to assess how well the simplified MOCCA-NoFB code
(which is very similar to the old version of the Monte Carlo code used
 previously to successfully simulate the evolution of real star clusters) can follow $N$-body results, and for which cluster properties it is enough to use the much faster and simplified code.

\subsubsection{Global parameters}\label{sec:global}

\begin{figure}
{\includegraphics[height=11cm,angle=270,width=9cm]{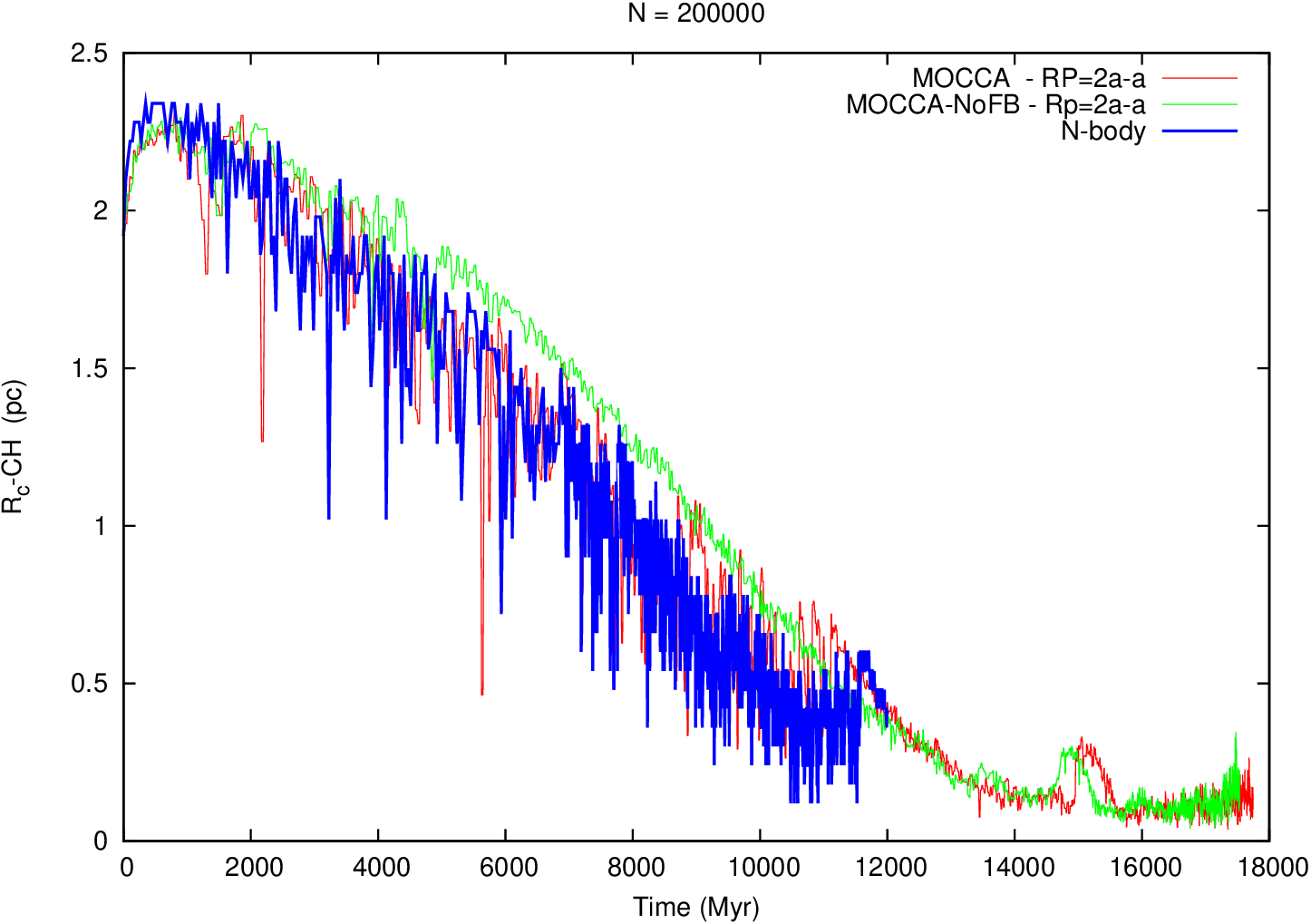}}
    \caption{Evolution of the core radius (defined according to \citet{CH1985}) for $N=200000$. Red line - MOCCA, green line - MOCCA-NoFB and blue line - $N$-body.}
\label{fig:rc}
\end{figure}

The comparison between $N$-body and MOCCA results was partially
discussed already in Secs.\ref{sec:scaling} and \ref{sec:free} for the
total mass evolution. It was shown that the agreement between the two
techniques is very good;  only for $N=24000$ was it less satisfactory,
at least with the globally preferred values of the escape
parameters (see Fig.\ref{fig:Mytid-m67}). The evolution of the core
radius (defined according to \citet{CH1985}) for $N=200000$ is shown
in  Fig.\ref{fig:rc}. The agreement between MOCCA and $N$-body is very
good. As one can expect, MOCCA-NoFB gives  slightly too large a core
radius. This, as  was explained in Sec.\ref{sec:free}, is connected
with the overestimation of binary energy generation in the cross
section regime for an excessively large value of $r_{pmax}$. The
large fluctuations in the core radius visible in the figure for the
$N$-body and MOCCA simulations are connected with the movement of a
massive BH or BH-BH binary in the system. The mass of such an object
is about $30-50M_{\odot}$. Its movement in the system is connected
with kicks acquired in interactions. If the massive object is present
in the core the core radius is smaller than when it is in the
halo. The mass of the massive object is comparable to the core
mass. When all the most massive objects are removed from the system,
because of strong interactions with other massive binaries or stars,
the evolution of the core radius is once again ``smooth". 
The evolution of the core radius is generally similar for the other
models (but without  large fluctuations) with the exception of
$N=100000$, for which the $N$-body results are systematically slightly
below those of MOCCA (see Fig. \ref{fig:Rc_kfallb}). 

\begin{figure}
{\includegraphics[height=11cm,angle=270,width=9cm]{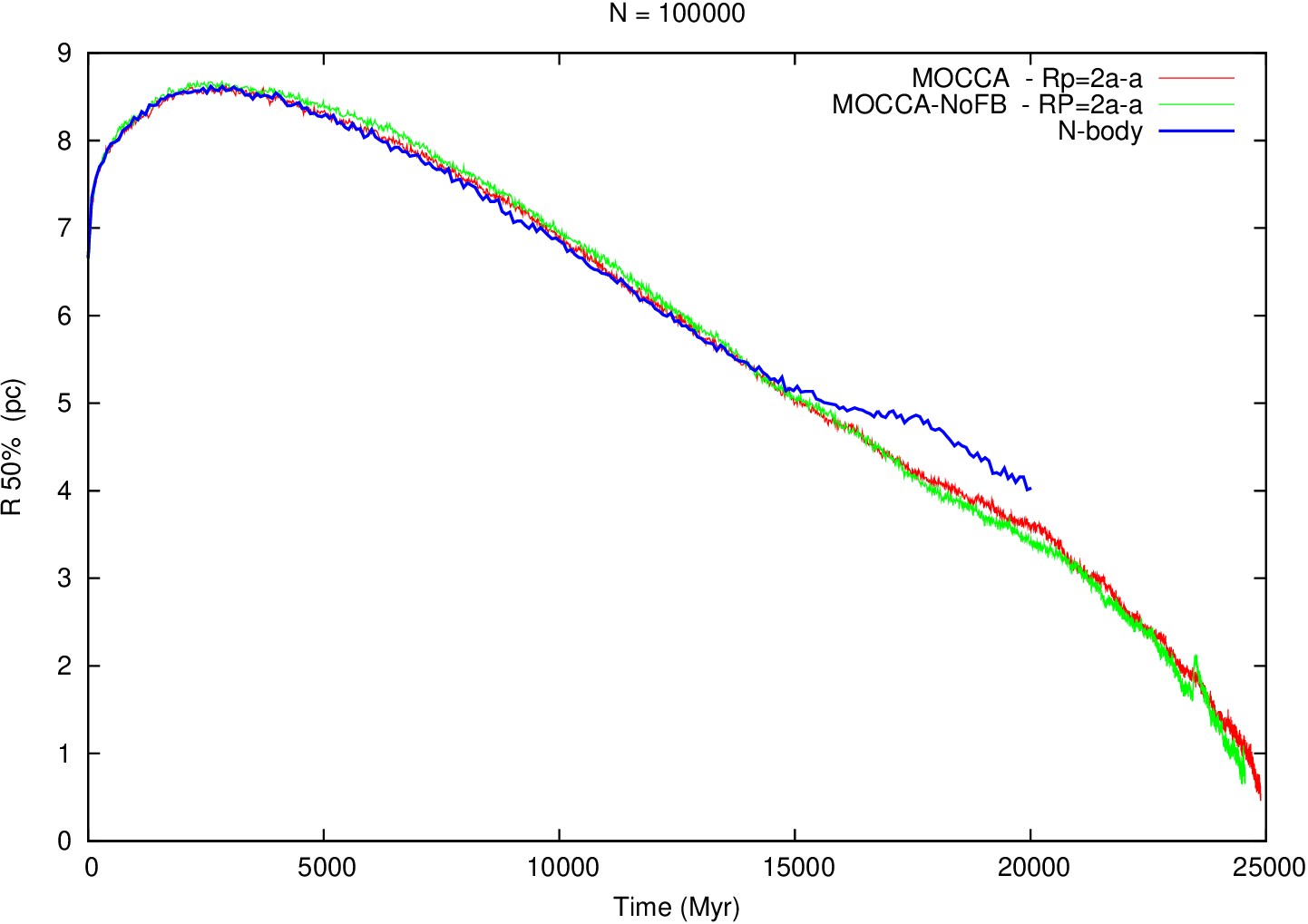}}
    \caption{Evolution of the half mass radius for $N=100000$. Red line - MOCCA, green line - MOCCA-NoFB and blue line - $N$-body.}
\label{fig:rh}
\end{figure}

The evolution of the half mass radius for the $N=100000$ model is
shown in Fig.\ref{fig:rh}. The evolution of all models  is very
similar from the very beginning, although the $N$-body results are
slightly below those of both MOCCA models. The differences start to build up
around the core collapse time, which seems to be around 17 Gyr for the
$N$-body model (and is visible as a  distinct
bump). For MOCCA models the core collapse time is around 20 Gyr. The
bump in the half mass radius is also visible for the MOCCA results,
but is less pronounced. That again suggests (see Sec. \ref{sec:half})
that the probability of binary formation is slightly too small in
MOCCA compared to $N$-body. For $N=24000$ and $N=200000$ the half mass
radius for the MOCCA models is systematically slightly below the
results of the $N$-body models. The differences start to build up from
the beginning, and are biggest around the time when the stellar
evolution stops  being the dominant process of cluster expansion
(i.e. the time when the indirect heating connected with  stellar mass
loss becomes smaller than the heating connected with binary energy
generation).  Then the evolution of the half mass radius starts to
converge for both $N$-body and MOCCA models. The same behaviour can be
observed for other Lagrangian radii (1\% and 10\%). The reason for
such  behaviour is unclear but may be attributable to the different
relative  strengths of the physical processes which operate during the
different phases of cluster evolution.  If mass segregation is faster
in $N$-body simulations than in MOCCA (evidence for which is given below), 
a more extended cluster could
be generated.  Stellar evolution, which is responsible for the loss of
stellar mass, can cause substantial expansion of the cluster, particularly in
the initial phases of evolution.  But both $N$-body and MOCCA models
rely on the same stellar evolution prescription
\citep[][]{Hu2000,Hu2002}, and so we cannot expect that the amount of
mass loss is different in the two models. However,  mass segregation
acting together with stellar mass loss can substantially amplify the
expansion effect.  If the most massive stars lose their envelopes when
they are already mass segregated the effect on cluster expansion is
larger.  Finally, the larger binary energy generation in $N$-body
simulations (see Figs.\ref{fig:ebin_50_t} and
  \ref{fig:ebin_10_50}, which show evidence of a larger average binary
  binding energy), or larger probability of binary formation , may be responsible for the faster expansion of Lagrangian radii.  
\begin{figure}
{\includegraphics[height=11cm,angle=270,width=9cm]{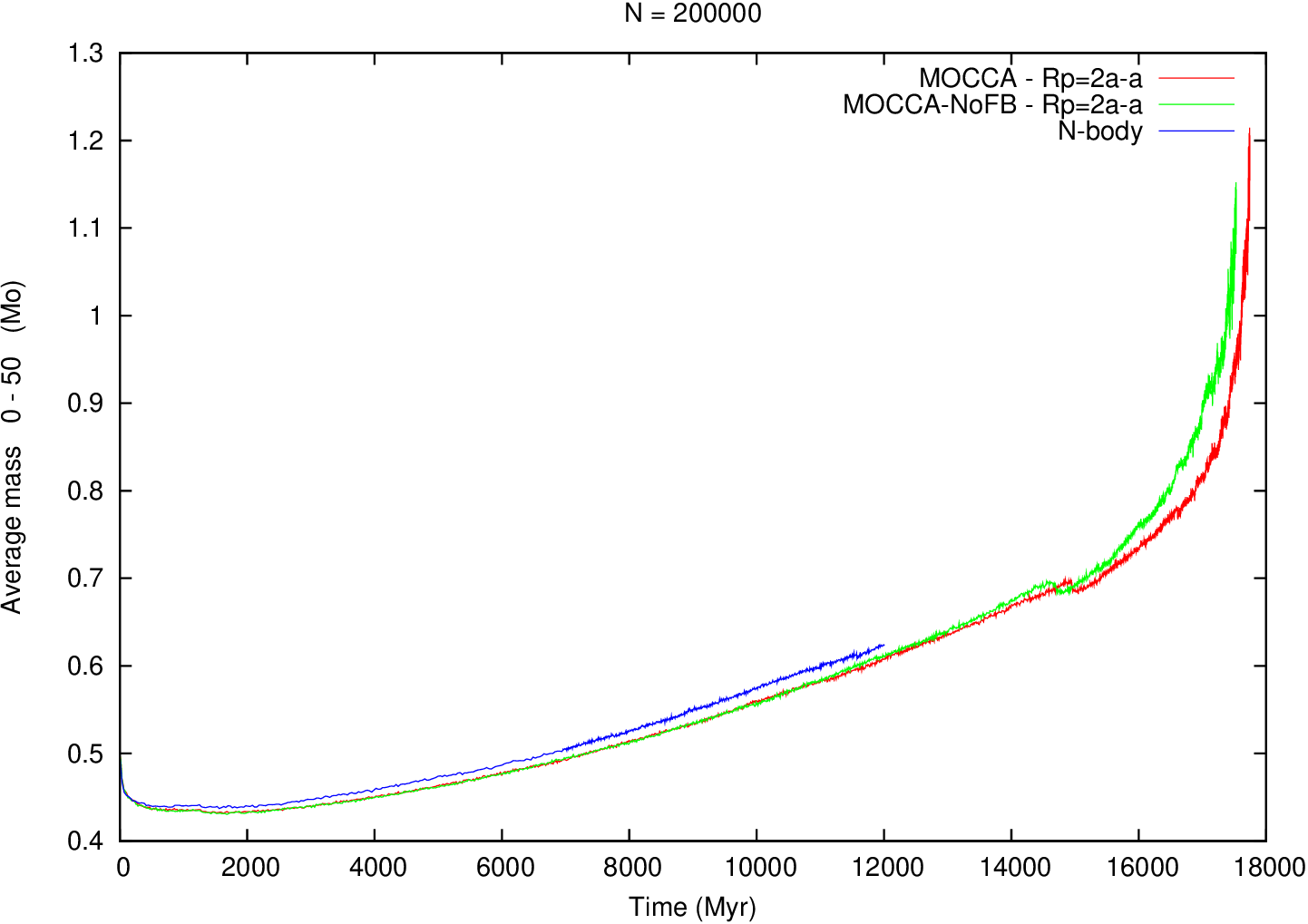}}
    \caption{Evolution of the average mass inside the 50\% Lagrangian radius for $N=200000$. Red line - MOCCA, green line - MOCCA-NoFB and blue line - $N$-body.}
\label{fig:am50}
\end{figure}
\begin{figure}
{\includegraphics[height=11cm,angle=270,width=9cm]{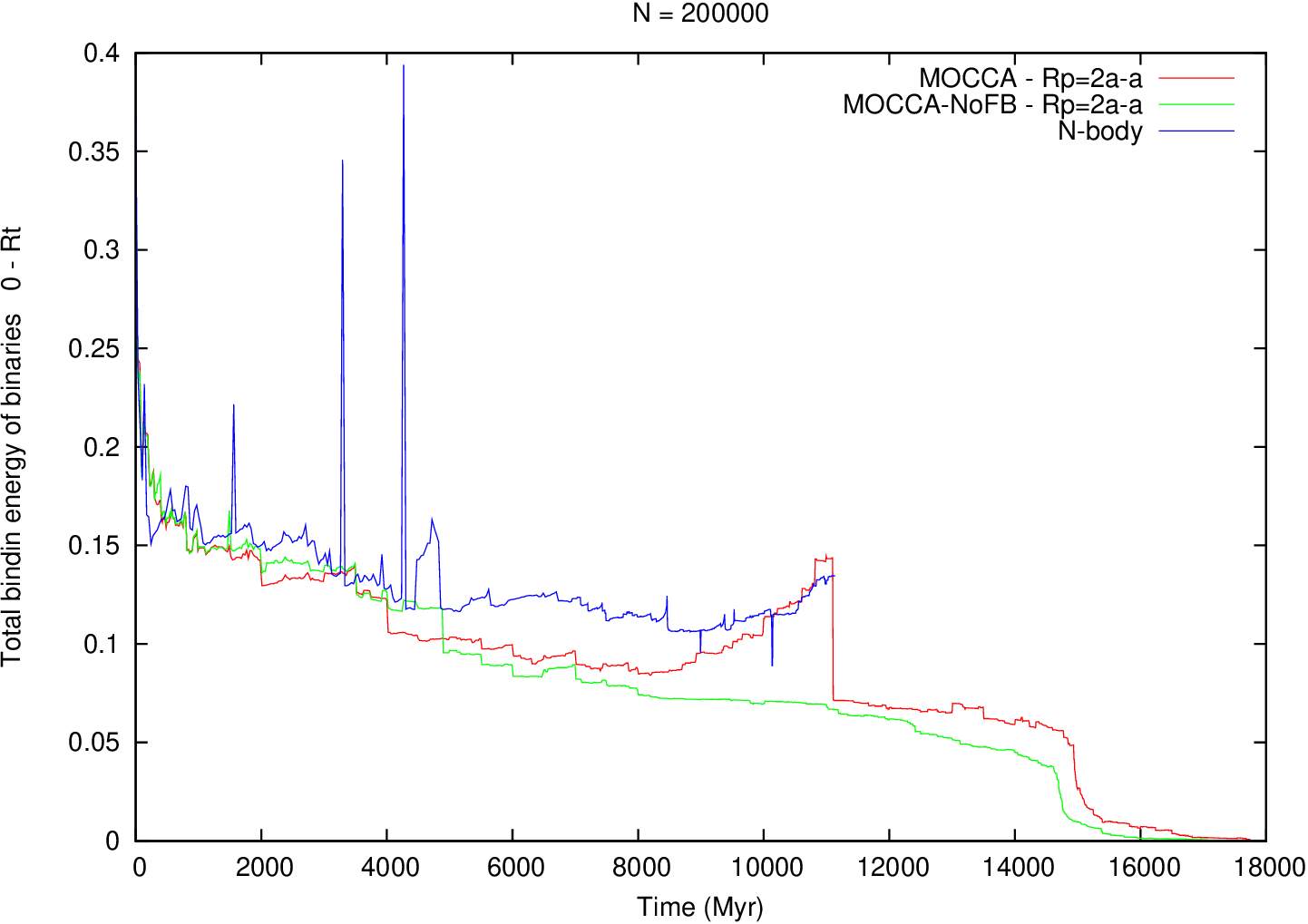}}
    \caption{Evolution of the total binding energy of binaries for $N=200000$. Red line - MOCCA, green line - MOCCA-NoFB and blue line - $N$-body.}
\label{fig:ebin}
\end{figure}

The evolution of the average mass inside the 50\% Lagrangian radius
and the evolution of the total binary binding energy for $N=200000$
are shown in Figs.\ref{fig:am50} and \ref{fig:ebin}. These figures are
representative for all models and Lagrangian radii.  It seems that
mass segregation is indeed slightly stronger in the $N$-body model
than in MOCCA. This suggests that  faster mass segregation in the
$N$-body model could be responsible, at least partially, for the
slightly discrepant evolution of the Lagrangian radii. The evolution
of the total binary binding energy is  rather similar from the very
beginning, until a later time (about 4 Gyr), when more energetic
binaries are formed in the $N$-body model (see also
Figs.\ref{fig:ebin_50_t} and \ref{fig:ebin_10_50}). The final
formation of a very hard binary is visible in the MOCCA and $N$-body
runs in Fig.\ref{fig:ebin} (see the discussion in
Sec.\ref{sec:binary}), but not for MOCCA-NoFB.  When this binary is
removed from the system, because of interactions, all models  are
similar once again. It is worth  noting that the results of simulations for MOCCA and MOCCA-NoFB are very similar until the late phases of evolution.

\subsubsection{Binary properties}\label{sec:binary}

The evolution of the number of binaries was already presented in
Sec.\ref{sec:free} during the discussion about the determination of
the free model parameters. We know that the evolution of the total
number of binaries is slightly too slow in MOCCA simulations. The
difference starts to build up at the time when stellar evolution
ceases to be the dominant process of cluster expansion (see
Fig.\ref{fig:Nb-rp-100k}).  

Now the average binary mass and  binding energy distributions will
be discussed for the model with $N=200000$.  The results for this
model are representative of the other models, but additionally it shows  a
buildup of the average binary mass and the binding energy of massive
binaries. This buildup is possible only for the model $N=200000$, for
which the distribution of supernova kicks is uniform between $0$
and $100$ km/s and allows a larger  fraction of BH to be bound to the
system than in the models with N=24000 and N=100000, which adopt a Maxwellian distribution with $\sigma=190 km/s$ for SN kicks.

\begin{figure}
{\includegraphics[height=11cm,angle=270,width=9cm]{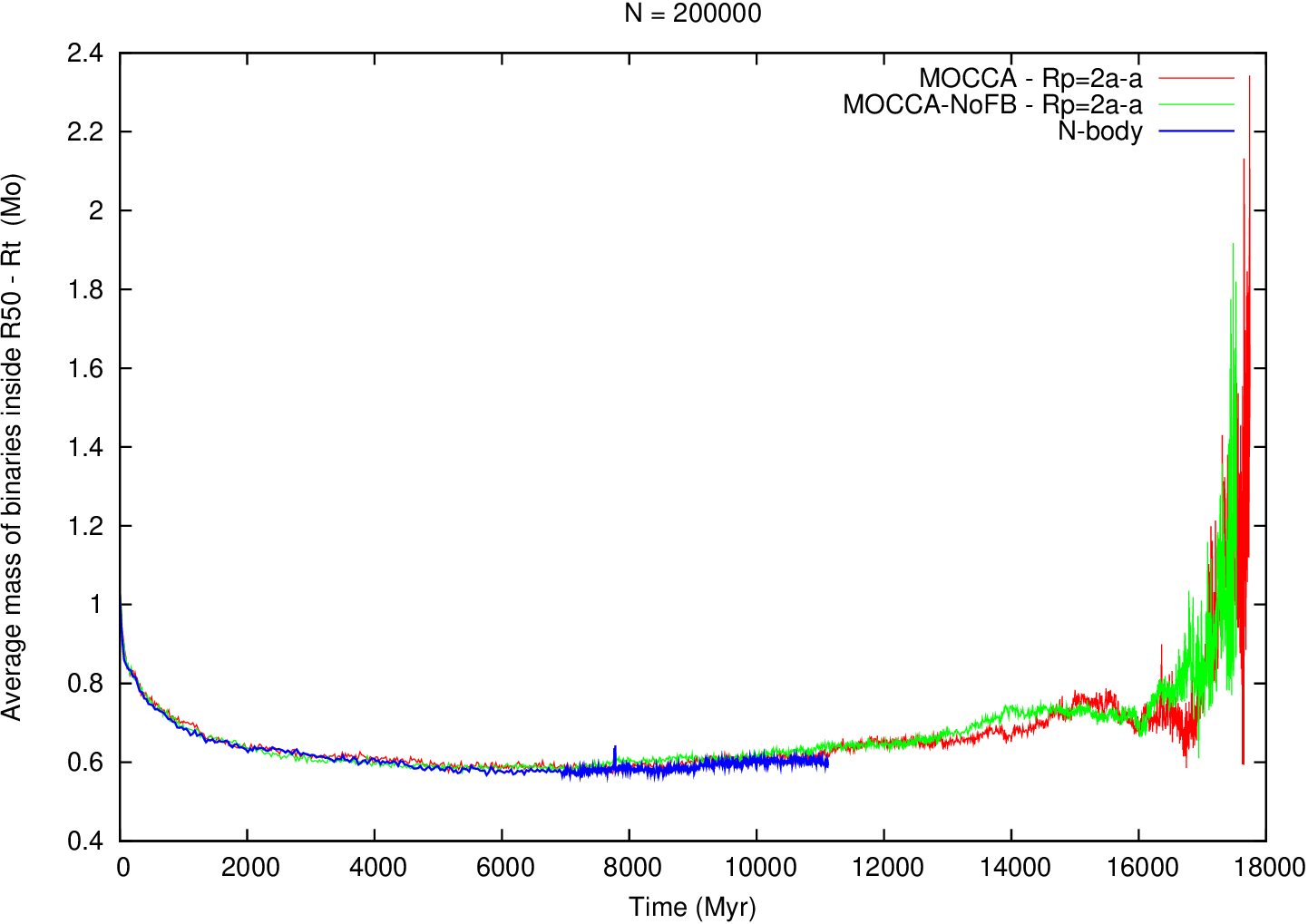}}
    \caption{Evolution of the average binary mass between the 50\% Lagrangian radius and the tidal radius for $N=200000$. Red line - MOCCA, green line - MOCCA-NoFB and blue line - $N$-body.}
\label{fig:Mbin_50_t}
\end{figure}

\begin{figure}
{\includegraphics[height=11cm,angle=270,width=9cm]{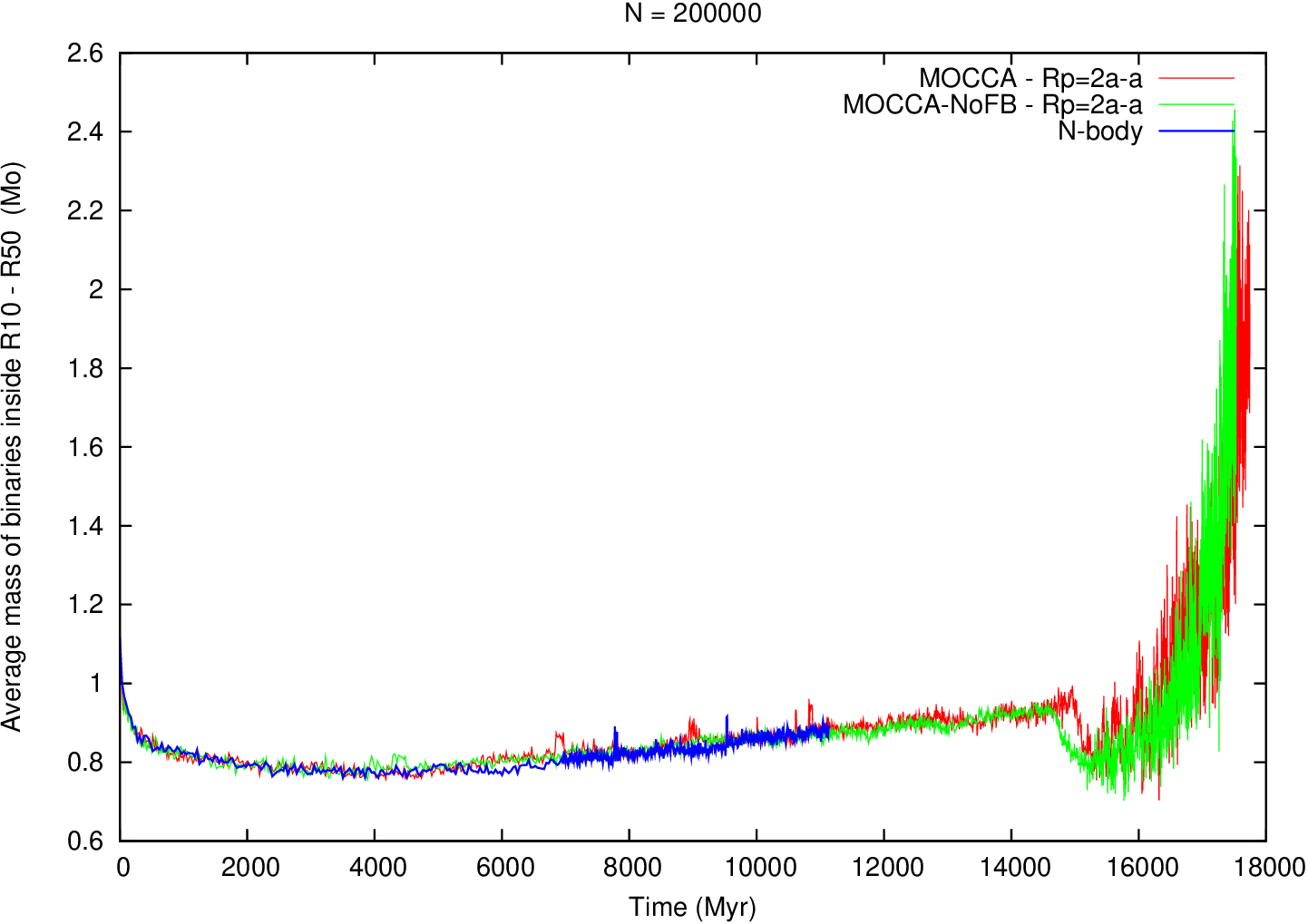}}
    \caption{Evolution of the average binary mass between the 10\% and 50\% Lagrangian radii for $N=200000$. Red line - MOCCA, green line - MOCCA-NoFB and blue line - $N$-body.}
\label{fig:Mbin_10_50}
\end{figure}

\begin{figure}
{\includegraphics[height=11cm,angle=270,width=9cm]{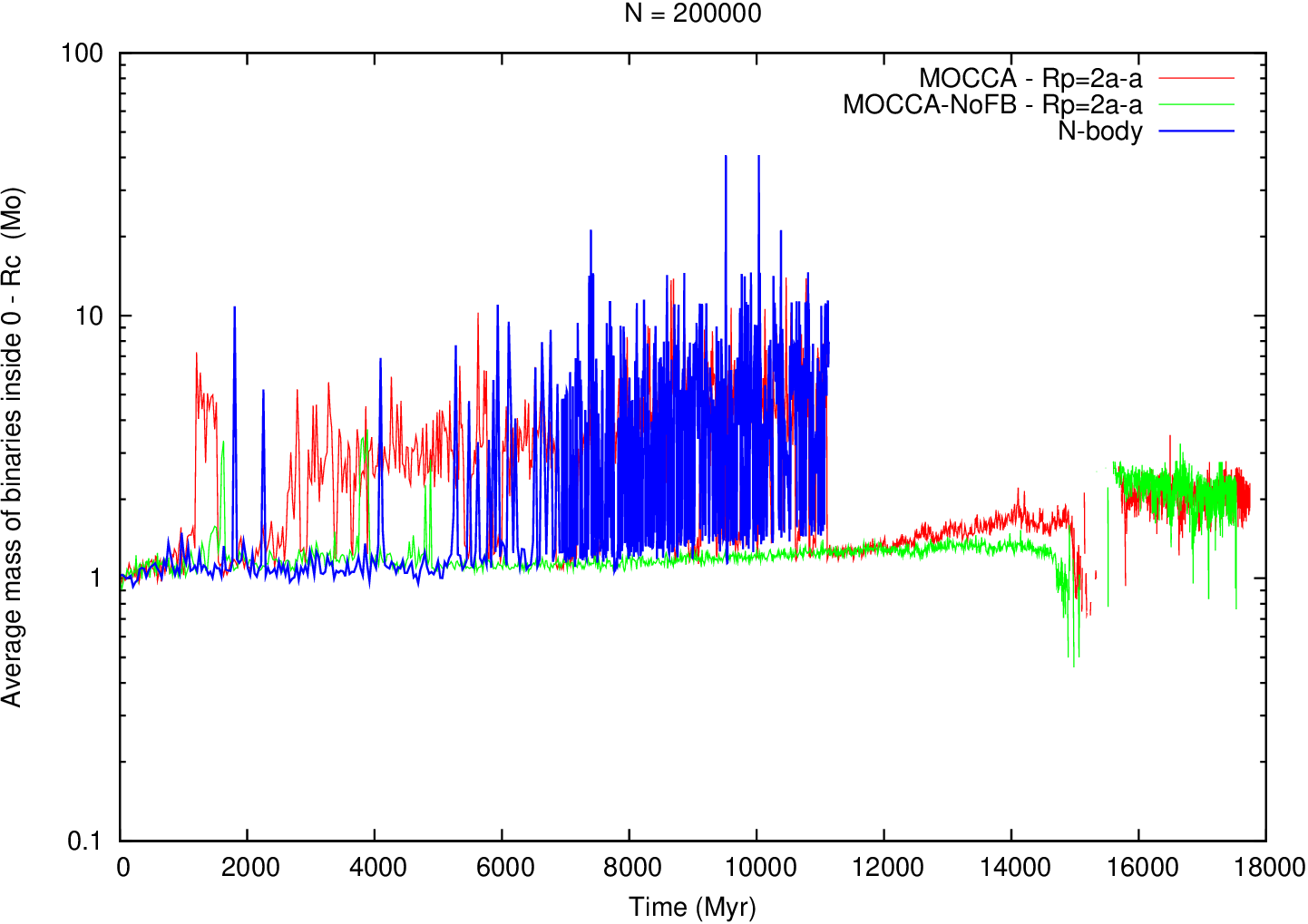}}
    \caption{Evolution of the average binary mass between the 0\% Lagrangian radius and the core radius for $N=200000$. Red line - MOCCA, green line - MOCCA-NoFB and blue line - $N$-body.}
\label{fig:Mbin_0_rc}
\end{figure}

The evolution of the average binary mass for different regions of the
system is shown in Figs.\ref{fig:Mbin_50_t}, \ref{fig:Mbin_10_50} and
\ref{fig:Mbin_0_rc}. The agreement between $N$-body and MOCCA results
is very good in all regions. This is despite the fact that, in the
$N$-body model, there is a smaller number of binaries than in MOCCA and
the Lagrangian radii are slightly different.  It seems that the
average binary mass and its distribution does not depend on the number
of binaries, which is probably a result of using exactly the same
binary initial conditions for both models and presumably a
similar mass spectrum of removed or destroyed binaries in both models.
For the region inside the core, the buildup of the mass of binaries is clearly visible. The big fluctuations are connected with the movement of the massive binaries, which because of hard interactions with other stars are kicked out from the core on very elongated orbits. 
Finally, they are kicked out of the system and the average mass of
binaries starts to become less chaotic and its changes become rather
small (see Fig. 7 in \citet{HS2012} for the $N$-body simulation).  The drop and then increase of the average mass around 14.5Gyr is connected with core collapse, but the $N$-body model stops before this occurs. 
The high central density makes binary-binary interactions very
effective, and a substantial number of relatively wide and massive
binaries are destroyed, causing a drop in the binary average mass.
This drop is visible well outside the core, presumably because of binaries on elongated orbits which occasionally visit the high-density part of the system. 
Looking at the average mass in different regions in the cluster we
clearly see evidence of mass segregation. In the centre the mean mass is about twice as large as in the halo. For the $N=200000$ model the difference between MOCCA and MOCCA-NoFB is very small except in the core, where the increase of the average mass is less pronounced for MOCCA-NoFB. The sharp increase of the average mass close to the cluster dissolution time is connected with the fact that only the most massive binaries are able to stay in the system; less massive ones are successively removed. 

In the models for other $N$ the behaviour of the average binary mass
is similar to that described above, except that in the core there
are no large fluctuations (as there are no very massive binaries
there), and for the late phases of cluster evolution small
discrepancies between the $N$-body and MOCCA models start to show up.

\begin{figure}
{\includegraphics[height=11cm,angle=270,width=9cm]{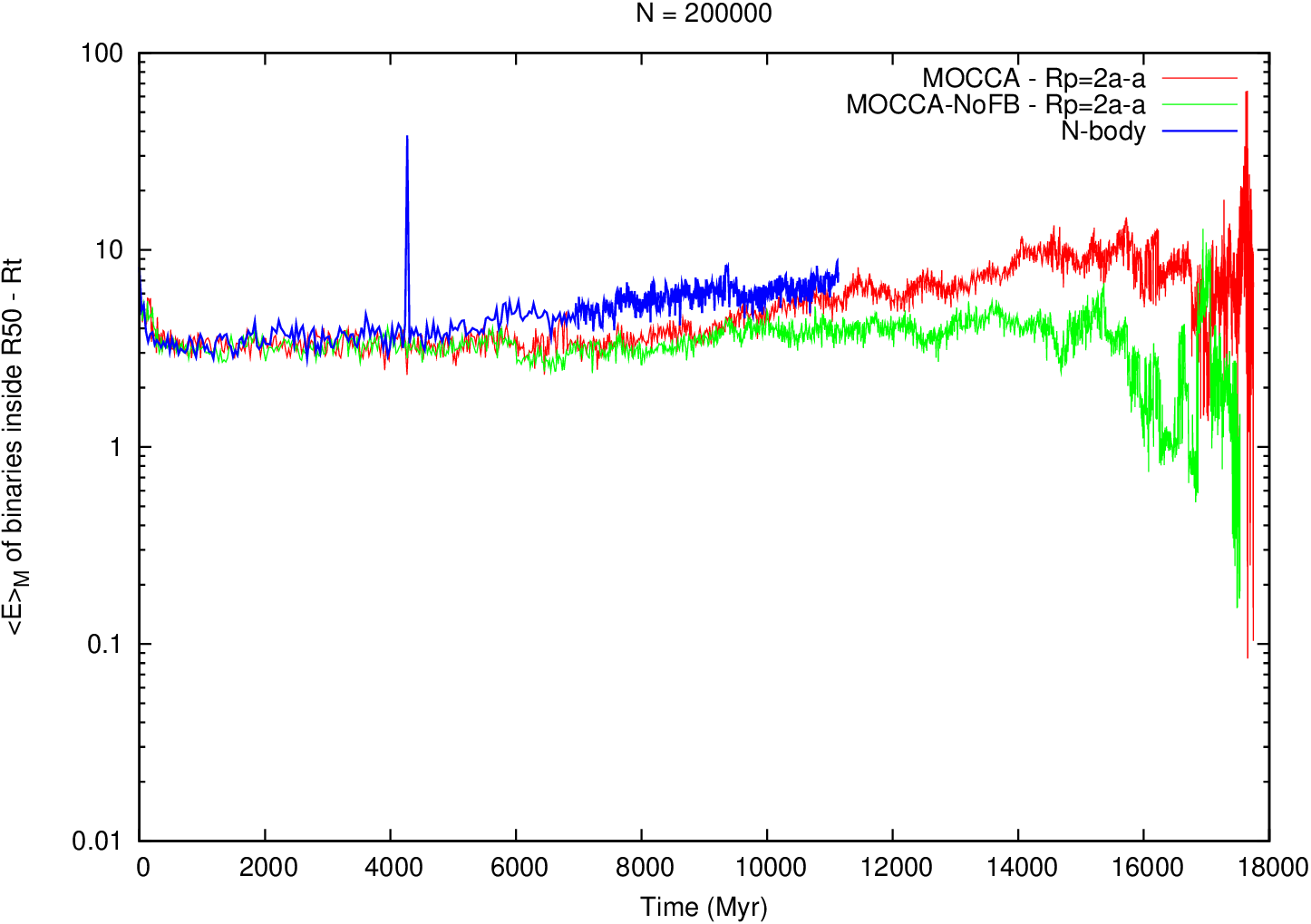}}
    \caption{Evolution of the average binary binding energy between
      the 50\% Lagrangian radius and the tidal radius for $N=200000$. Red line - MOCCA, green line - MOCCA-NoFB and blue line - $N$-body.}
\label{fig:ebin_50_t}
\end{figure}

\begin{figure}
{\includegraphics[height=11cm,angle=270,width=9cm]{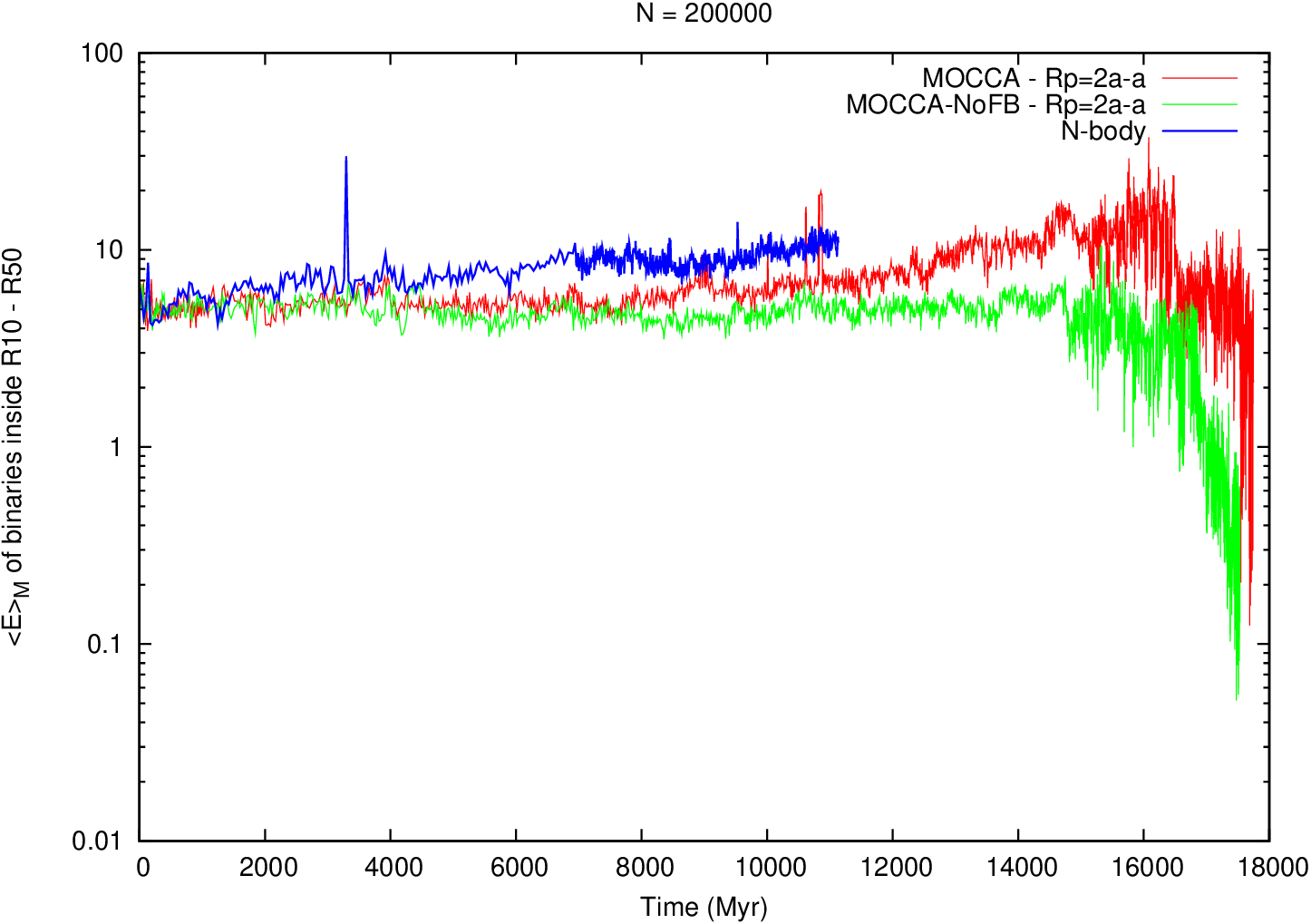}}
    \caption{Evolution of the average binary binding energy between
      the 10\% and 50\% Lagrangian radii for $N=200000$. Red line - MOCCA, green line - MOCCA-NoFB and blue line - $N$-body.}
\label{fig:ebin_10_50}
\end{figure}

\begin{figure}
{\includegraphics[height=11cm,angle=270,width=9cm]{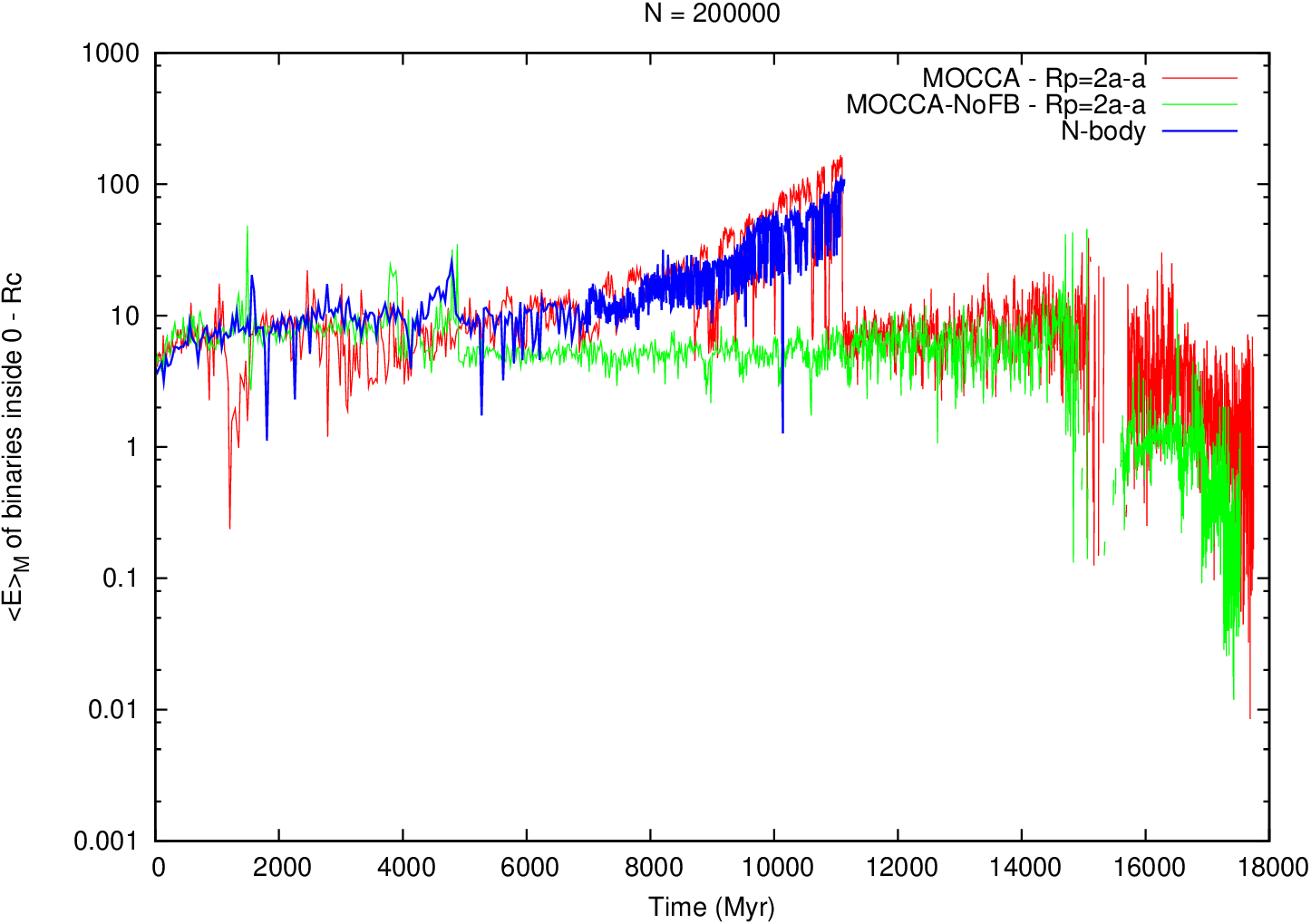}}
    \caption{Evolution of the average binary binding energy between
      the 0\% Lagrangian radius and the core radius for $N=200000$. Red line - MOCCA, green line - MOCCA-NoFB and blue line - $N$-body.}
\label{fig:ebin_0_rc}
\end{figure}

The evolution of the average binary binding energy for different
regions of the system is shown in Figs.\ref{fig:ebin_50_t},
\ref{fig:ebin_10_50} and \ref{fig:ebin_0_rc}. The best agreement
between the $N$-body and MOCCA results is for the core region. The
buildup of the binary binding energy connected with the formation of a
massive BH-BH binary is clearly visible. It has to be stressed that
the increase of the binary binding energy is not connected with core
collapse and core bounce. It is purely connected with the formation of
a very massive BH-BH binary and then the increase of its binding
energy in interactions with stars and other binaries. The binary is
finally removed from the system and then the average binary binding
energy suddenly drops. For other cluster regions the agreement between
the $N$-body and MOCCA results is less satisfactory; the average
binary binding energy for the $N$-body model is systematically larger
than for MOCCA. The differences start to build up just after the time
when stellar evolution ceases to dominate the  global evolution of the
cluster. It seems that in $N$-body simulations harder binaries can be
produced in the core before they are kicked out to the outer parts of
the system. Maybe this is connected with the fact that in MOCCA only binaries are allowed; higher hierarchies are not allowed, as triples and quadruples are artificially disrupted into binaries and single stars (see \citet{HyG2012}). It is well known that in $N$-body simulations substantial numbers of triples and quadruples are formed \citep[e.g.][and reference therein]{mikkola1984,MHM1991,HH2003,Hurleyetal2005}. They can interact with other objects in the system, and produce on average slightly harder binaries.
In contrast with  the average binary mass, the average binary binding
energy for MOCCA-NoFB is clearly smaller than for MOCCA and the
$N$-body results. In the MOCCA-NoFB simulations Heggie's cross section was
used (see Sec.\ref{sec:prob}). Therefore, on average the binary
binding energy changes by 40\% in every interaction. The
larger the changes of the binary binding energy due to interactions
are, the smaller the binary binding energy is at the time just
before escape, and so the average binary binding energy outside the
core is smallest for MOCCA-NoFB (see Figs.\ref{fig:ebin_50_t},
\ref{fig:ebin_10_50}). For the MOCCA simulations with $r_{pmax}=2a$
for binary-single interactions, the average change of binding energy
is about 18\%. It seems, from the point of view of the underabundance
of binaries with large enough energies in the outer parts of the
system, that this number is still too large.   Indeed, larger
$r_{pmax}$ gives much better agreement with the $N$-body results (the
larger $r_{pmax}$ the smaller the average change of binary binding
energy), but it is still not perfect. We should keep in mind that,
if we choose  too large a value of $r_{pmax}$ in order to get smaller
binding energy changes in interactions, we create a much larger number
of BSS compared to $N$-body simulations (see Sec.\ref{sec:prob} and
Fig. \ref{fig:Nbs-rp-100k}). In this paper we decided not to use
extremely large values of $r_{pmax}$, putting more emphasis on the
number of BSS than on the distributions of the average binary binding
energy. The dependence of the number of BSS on $r_{pmax}$ is much
stronger than for the binary binding energy. But if one is more
interested in the binary binding energy distribution, much larger
$r_{pmax}$, say $9a$ or even higher, should be used. It is worth
stressing that MOCCA-NoFB cannot reproduce the binary binding energy
distribution observed in the $N$-body and MOCCA results, except for the $N=24000$ models, where both models give very similar results.

For other $N$-body models the behaviour of the average binary binding
energy is similar to that described above, except that in the core
there is no large buildup of very hard binaries. For $N=24000$ the
discrepancies between the $N$-body model and the two MOCCA models
  are largest, except in the core, where all models agree. This is
probably connected with the fact that this model contains as much as
50\% binaries (10, 20 times more than in the other models), and small
differences in binary energy generation may produce larger
discrepancies in the properties of the binary distributions. Indeed, for $N=200000$ the discrepancies are smallest.

\subsubsection{Black holes and blue stragglers}\label{sec:bss_bh}

The $N$-body simulations used for the comparison with MOCCA provided
additional information about global system parameters and detailed
data about binaries and about BH and BSS. Some data on BSS was
already used in Sec.\ref{sec:free} to put some constraints on
$r_{pmax}$. In this section we will try to assess how well MOCCA can reproduce the $N$-body results with respect to the behaviour of ``peculiar" objects, which are rare and are formed in interaction channels which are not very probable.
\begin{figure}
{\includegraphics[height=11cm,angle=270,width=9cm]{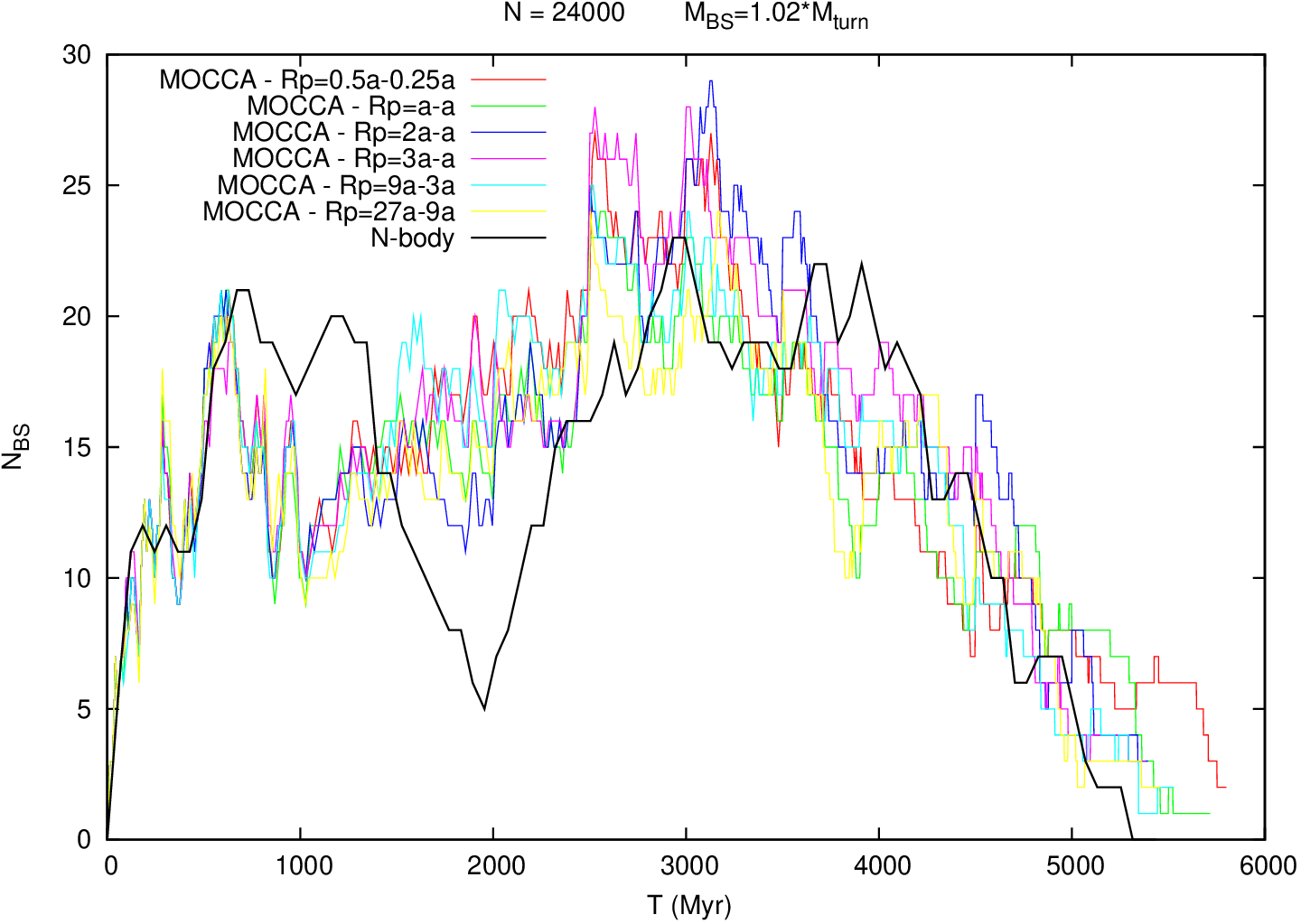}}
    \caption{The evolution of the number of blue stragglers in the system as a function of time for MOCCA models with different $r_{pmax}$ and $N$-body simulations for $N=24000$.}
\label{fig:Nbs-rp-m67}
\end{figure}     

In Fig.\ref{fig:Nbs-rp-m67} the evolution of BSS for different
$r_{pmax}$ for $N=24000$ is shown. The dependence of the number of BSS
on $r_{pmax}$ is different from what is found for $N=100000$ and
$N=200000$. There is no sharp increase of the maximum number of BSS
for $r_{pmax} > 3a-a$ (see Fig.\ref{fig:Nbs-rp-100k}). For $N=24000$
there is at most a very weak dependence of the number of BSS on
$r_{pmax}$. The fluctuations of the number of BSS connected with the
stochasticity of the system and MOCCA model are comparable in  size
with differences related to $r_{pmax}$ - about $5$ BSS. The
differences observed between Figs.\ref{fig:Nbs-rp-100k} and
\ref{fig:Nbs-rp-m67} are rather surprising, and point in the direction
of a strong dependence on the binary fraction. It was 50\% in
$N=24000$, at least ten times more than in the other models. This
result contradicts the naive idea that a large number of binaries
should amplify the effect of large $r_{pmax}$.  Perhaps a large binary
fraction makes binary-binary interactions more probable and so more
binaries are destroyed, or binary properties are changed in such a way
that the formation of BSS is less probable, somehow balancing the
larger BSS formation for larger $r_{pmax}$.  At any rate, MOCCA is
able to reproduce the evolution of the number of BSS given in $N$-body
simulations provided that a reasonable value of $r_{pmax}$ is chosen.
MOCCA-NoFB gives  a much smaller number of BSS than the other
models. That is not surprising, because channels which are important for BSS formation  are missing in the cross section approach. 

There is a long debate about BH kick velocities in SN explosions. We
tested three possible variants in the paper (guided by choices made in
Jarrod Hurley's $N$-body simulations).  The  first variant is a
Maxwellian distribution with $\sigma = 190$ km/s \citep{Ph1992}, or a
Maxwellian distribution with $\sigma = 190$ km/s but with the maximum
kick velocity set to $50$ km/s (for the $N=24000$ and $N=100000$ models). Such a reduction of the kick velocity to a maximum of $50$ km/s is the procedure used in Jarrod Hurley's $N$-body simulation. These variants are referred to as $kick = 0$ and $kick = 0-r$, respectively. The second variant is a uniform
distribution with kick velocities between $0$ and $100$ km/s, or a
uniform distribution with kick velocities between $0$ and $50$ km/s,
referred to as $kick = 1$
  and $kick = 1-r$, respectively (but
only for the $N=200000$ model). The third variant is a Maxwellian
distribution with $\sigma = 190$ km/s, but the kick velocity is
finally modified by the amount of mass which falls back on the
BH during the SN explosion \citep{Belczynskietal2002}; referred to as
$kick=2$.  The amount of mass fall-back depends on the core mass of a star just before the supernova explosion. For a core mass greater than $7.6M_{\odot}$ the kick velocity is $0$ km/s. 

Discussion of the way in which the different assumptions about SN
kicks influence both the evolution  of the system  and the properties
of BH and BH-BH binaries will be presented for the $N=100000$ model.
For other models the different kick assumptions  do not have a strong
influence on the system parameters.  For $N=24000$ the escape velocity
is very small, and so even for the $kick=2$ case only a few BH will stay in the system after SN kicks,  and they will be quickly
removed by interactions within the system.  For $N=200000$ the SN kick
velocities for the $kick=1$ and $kick=1-r$ cases were already relatively small, and a substantial fraction of BH were left in the system.  Therefore in the case  $kick=2$ the BH fraction did not change substantially and the system evolution was not influenced strongly.  For $N=100000$, on the other hand, the situation is
different.  Only in the case $kick=2$ was a substantial number of BH left in the system, and these strongly influence its evolution
in a similar way to the case $N=200000$.
\begin{figure}
{\includegraphics[height=11cm,angle=270,width=9cm]{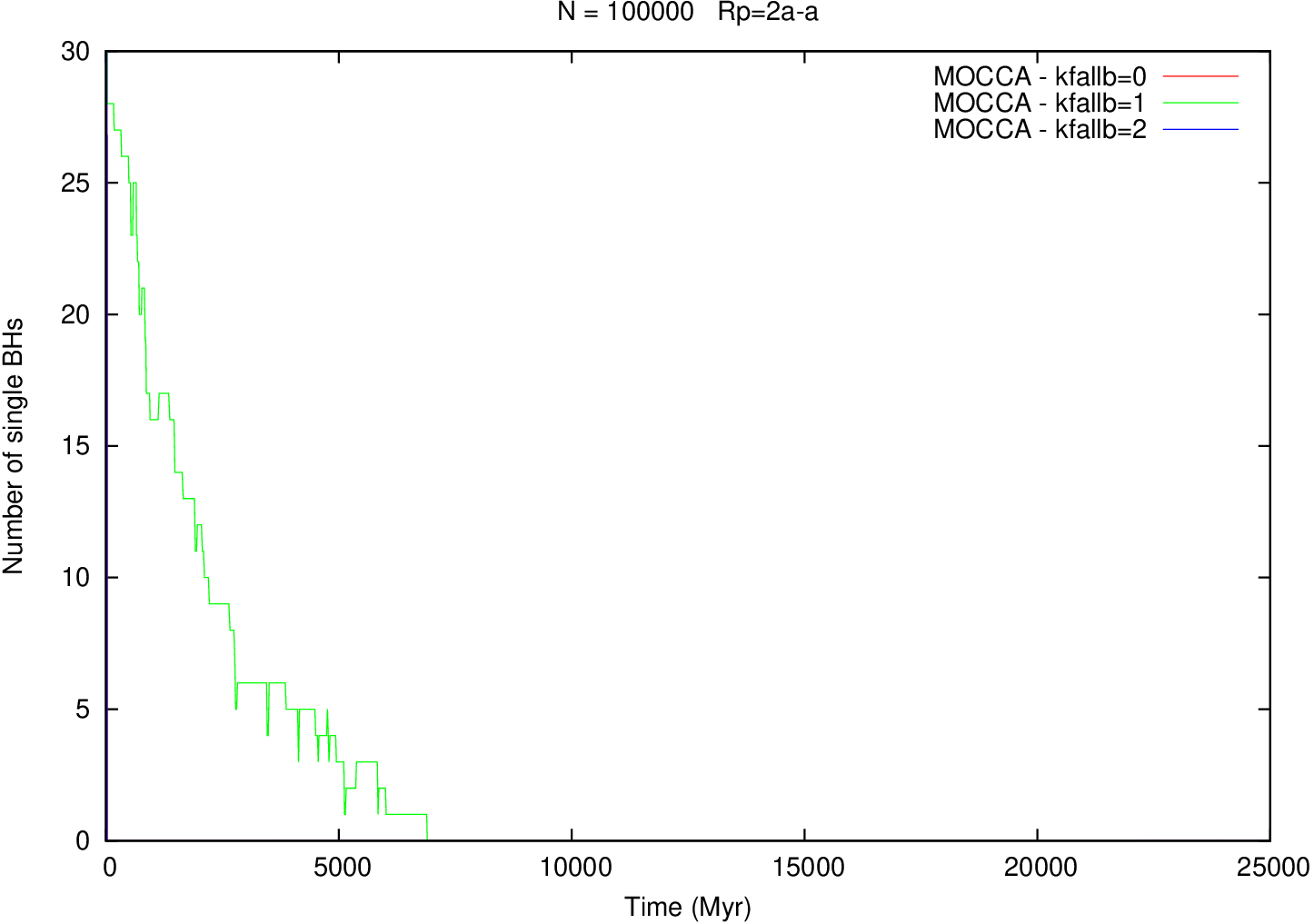}}
    \caption{Evolution of the number of BH in the system for $N=100000$ and  $kick=2$. For cases $kick=0$ and $ 0-r$ practically all BH escaped from the system just after their formation. See description of the $kick$ parameter in the text.}
\label{fig:N_BH_kfallb}
\end{figure}
\begin{figure}
{\includegraphics[height=11cm,angle=270,width=9cm]{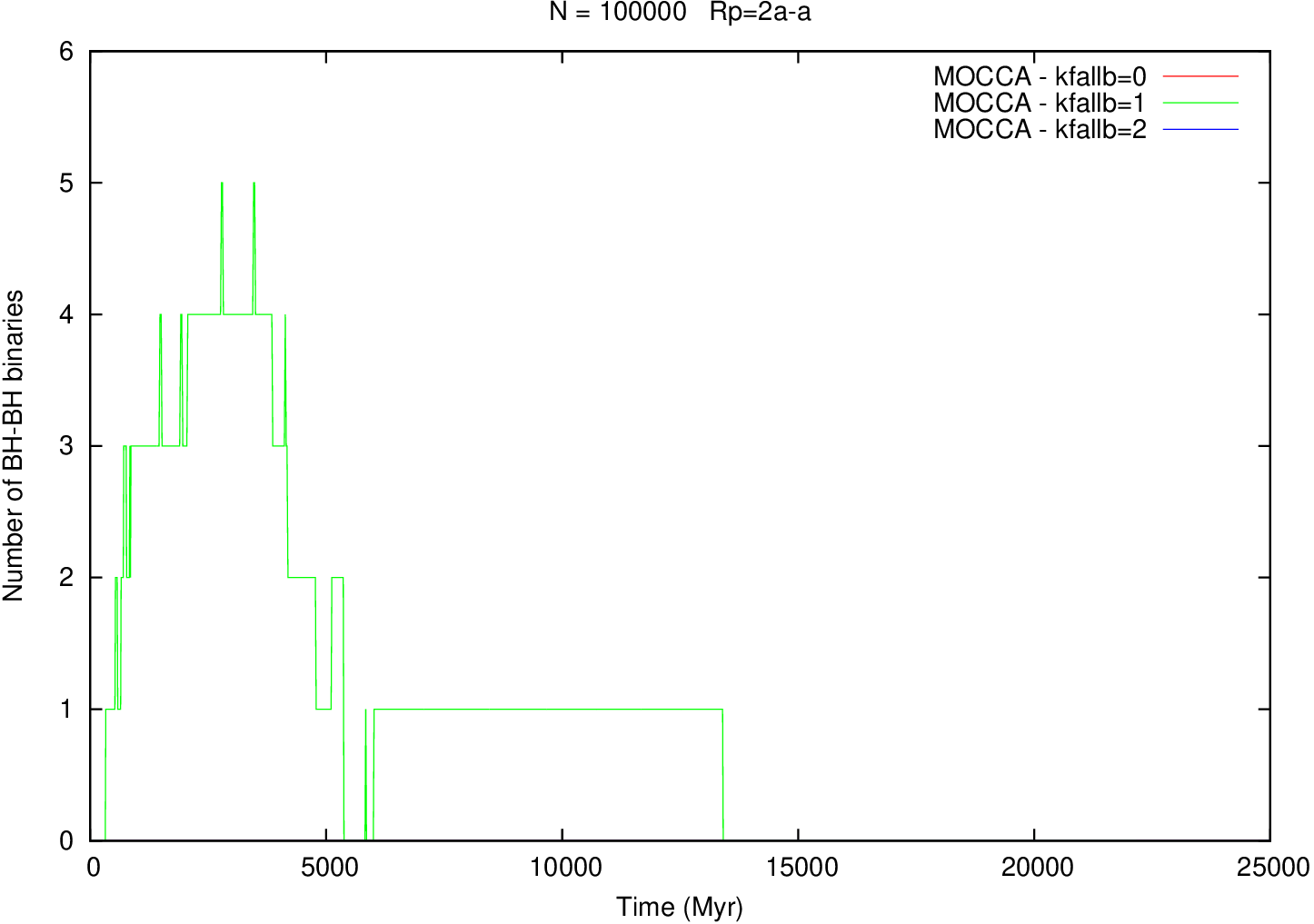}}
    \caption{Evolution of the number of BH-BH binaries in the system
      for $N=100000$ and $kick=2$. For cases
      $kick=0$ and $0-r$}
      there are no BH-BH binaries because no BH are left after the SN kicks. See description of the $kick$ parameter in the text.
\label{fig:N_BH-BH_kfallb}
\end{figure}

In Figs.\ref{fig:N_BH_kfallb} and \ref{fig:N_BH-BH_kfallb} the number
of BH and BH-BH binaries are shown for $N=100000$ and for different
assumptions about the SN kick velocities.  The striking feature of
these figures is the fact that there are no BH and BH-BH binaries
left in the system after SN kicks for case $kick = 0$ and even for
$kick = 0-r$.  Kick velocities are too large and all BH escape immediately from the system.  Only in the case $kick = 2$ were kick velocities  small enough to keep a substantial number of BH in the system.  The most massive BH have masses larger than $25M_{\odot}$,
and so  in three-body interactions they very quickly form massive
binaries with mass about $50M_{\odot}$.  The probability of binary
formation in  three-body interactions is a very strong function of
masses \citep{He1975,Gi2001}.  As can be seen from
Fig.\ref{fig:N_BH-BH_kfallb} the number of BH binaries present at the
any one time in the system can be as large as 5. They are the most
massive objects in the system, but this does not mean that all of
these binaries are  in the core at the same time.  They strongly
interact with other stars/BH or binaries and acquire (dynamical)
kicks which move them outside the core.  When they become hard enough
they are  removed from the system one by one in a final strong
interaction.  In the course of interactions each massive BH-BH binary
can remove several stars, including a few BH.  They are acting like a
vacuum cleaner quickly reducing the number of single BH in the
system, which is clearly visible in Figs.\ref{fig:N_BH_kfallb} and
\ref{fig:N_BH-BH_kfallb}. As has been mentioned, the
probability of a binary-single interaction depends on the mass of the
interacting objects: the larger the mass the larger the probability,
and the larger the number of hard interactions resulting in
substantial energy generation.  The hardest and most massive binaries
can produce much more energy in interactions than other binaries.
Therefore, one can expect that such a large amount of energy is able
to change the system structure, resulting in distinct features not
visible in models with $kick=0$ and $kick=0-r$. The following paragraphs confirm this expectation.

\begin{figure}
{\includegraphics[height=11cm,angle=270,width=9cm]{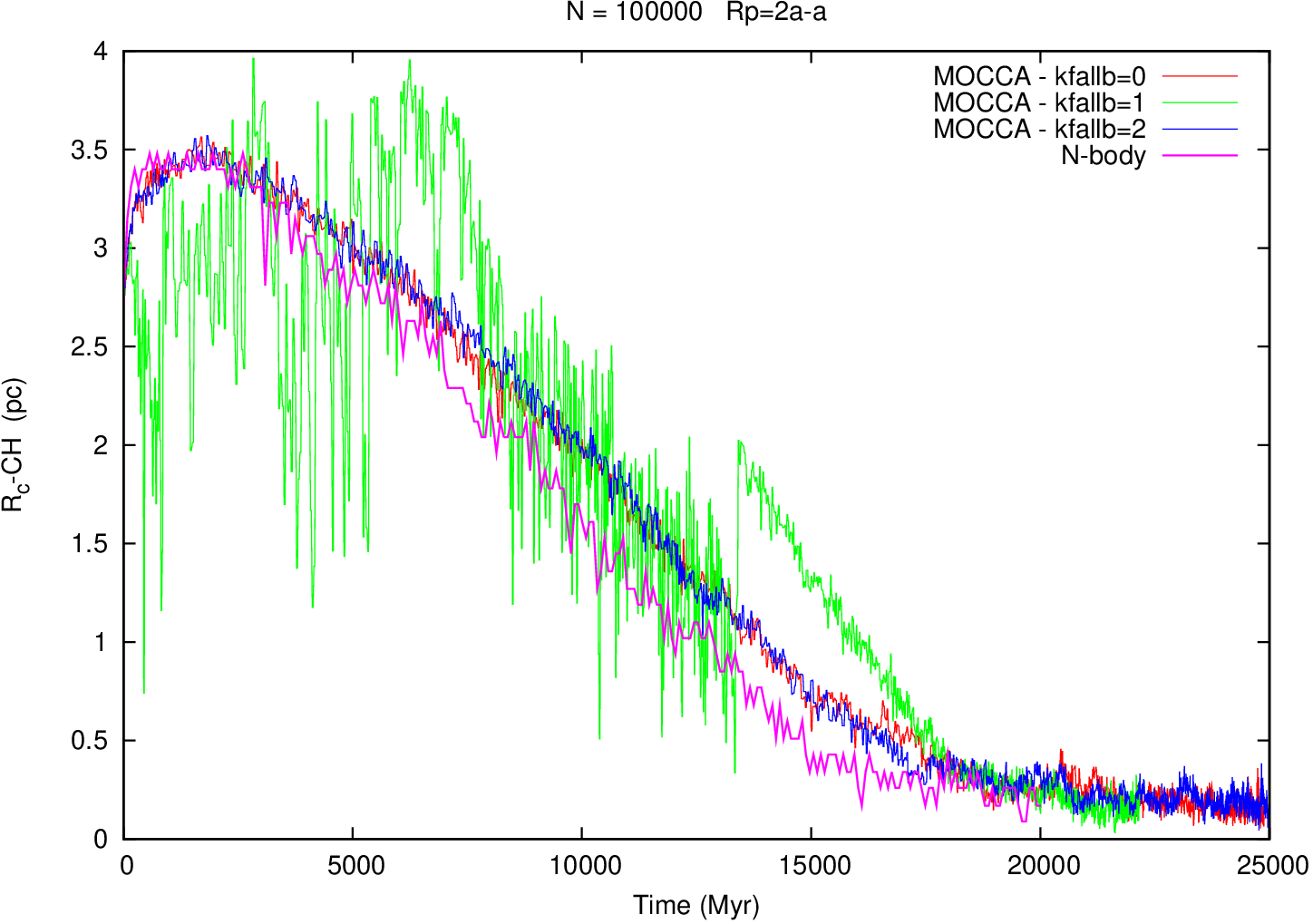}}
    \caption{Evolution of the core radius for $N=100000$. Red line - $kick = 0$, green line - $kick = 2$ and blue line - $kick = 0-r$. See description of the $kick$ parameter in the text.}
\label{fig:Rc_kfallb}
\end{figure}
\begin{figure}
{\includegraphics[height=11cm,angle=270,width=9cm]{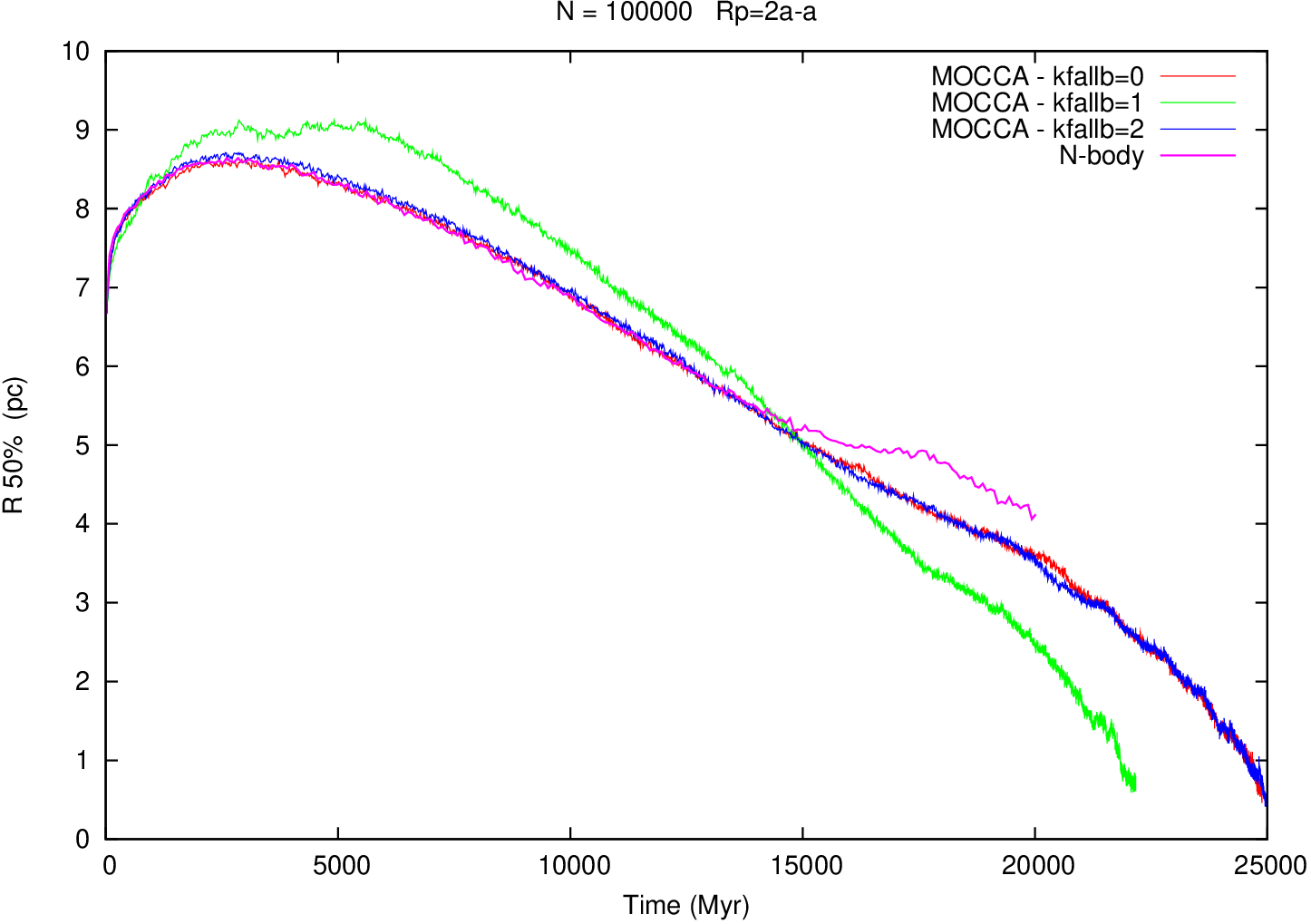}}
    \caption{Evolution of the half mass radius for $N=100000$. Red line - $kick = 0$, green line - $kick = 2$ and blue line - $kick = 0-r$. See description of the $kick$ parameter in the text.}
\label{fig:Rh_kfallb}
\end{figure}
Figs.\ref{fig:Rc_kfallb} and \ref{fig:Rh_kfallb} show the evolution of
the core and half mass radii for different $kicks$ and the $N$-body model for $N=100000$.  It can be seen that, in the case when massive
BH and massive BH-BH binaries are present in the system, the
evolution of the core radius is very stochastic, characterized by
large and fast fluctuations, on the time scale of order  several dozen
Myr.  Also, the half mass radius in this case evolves faster and has
larger values, which is presumably connected with stronger energy
generation.  One consequence of larger energy generation is much
faster system evolution: the cluster has a  shorter lifetime.
Therefore, it is not only BH with masses of several hundred
$M_{\odot}$ (IMBH), which can influence the system structure, that can
be detected by observations; also the presence in the system of
several massive stellar-mass BH can leave an observational imprint on
the cluster structure.  That is a very interesting possibility which
is worth checking further.  Similar behaviour was reported in
\citet{Hu2007}, i.e. stochastic evolution of the core radius powered
by a single rather massive BH, and in \citet{Merrittetal2004} and
\citet{Mackeyetal2008}, which studied strong core expansion in a
massive star cluster powered by a large number of stellar mass BH; in
that case, however, the evolution of the core radius was rather
smooth,  because of the large number of BH,  which form a bound subsystem in the central part of the cluster.

The behaviour of the half mass and core radii described above for
$N=100000$ is also visible for $N=200000$.  The fact that, in this
case, the numbers of BH and BH-BH binaries are very similar, for
different $kicks$, is interesting.  This suggests that it is not the number of BH-BH binaries that is crucial for the  behaviour discussed
above, but the mass of the BH-BH binaries.  For the case $kick=2$, more massive BH can be formed than in other $kicks$, because of the mass fallback during SN explosion \citep{Belczynskietal2002}.
Therefore, the most massive binary which can form in this case is
substantially more massive than in other cases.  Such a binary can
generate more energy in interactions, and remove more stars and
binaries from the system, than a less massive binary.  It is possible
that in the case when two or more massive BH-BH binaries are formed in
the system, they will interact between themselves and quickly harden.
Then there is a non-zero probability that, before they can escape,
they will merge in collision interactions with ``ordinary"
stars/binaries and form a more massive BH, which will very quickly
again form a binary. 
The sequence can be repeated a few times forming a seed IMBH in the
system.  This is an interesting new road for the possible creation of
IMBH in globular clusters.  This possible scenario strongly relies on
the SN kick velocity distribution for BH, the initial mass function
for massive stars and the initial cluster concentration. 

It is worth stressing that the faster MOCCA-NoFB code can be used for
projects which investigate the evolution of star clusters from the
point of view their global properties.  If one is interested in
properties of "peculiar" objects and their distributions, however, the
slower MOCCA code with the Fewbody integrator should be used.

At the end of this section we would like to refer again to the
choice of the best value of $r_{pmax}$.  We argued in
Sec. \ref{sec:free} that the best compromise seems to be $r_{pmax} =
2a$ for binary-single dynamical interactions and $r_{pmax}=a$ for binary-binary interactions.  This means that we put more emphasis on the limitations given by BSS than those  given by binaries.  The reason for that is as follows.  The BSS definition in the $N$-body and MOCCA codes is
exactly the same, and the fluctuations in the number of BSS are
smaller than the number  itself.  Therefore the sharp increase in the maximum
number of BSS with larger $r_{pmax}$ (observed in
Fig.\ref{fig:Nbs-rp-100k}) is real, and may be understood as follows. Very distant fly-bys do not   change binary binding energies, but mainly increase binary   eccentricities.  Larger eccentricity means that binary components
can approach each other more closely, substantially increasing the
possibility of tidal interactions or mass transfer between   components,  and so leading to more abundant BSS formation.   Regarding the number of binaries in the system, there are some ambiguities  about the operational binary definition in the $N$-body   and MOCCA codes.  In an $N$-body simulation it is easy to count
regularised binaries (KS binaries), but it is not quite so
straightforward to find all non-KS binaries.  In MOCCA  many
binaries are directly followed which would not be KS binaries in
  an $N$-body simulation.  As we pointed out in Sec.\ref{sec:free}, however, the number of non KS binaries is rather small and cannot be entirely responsible for the observed differences in the evolution of the number of binaries.  For the evolution of the binary binding energy distribution, the evidence is rather indirect, and the observed dependence on $r_{pmax}$ may also be at least partially attributable to as yet unidentified physical processes (e.g. the influence of close neighbours on the binary interaction process) or some systematic errors in MOCCA.  Therefore,  we decided to opt for the compromise value of $r_{pmax}$ rather than use the larger values suggested by the total number of binaries and their
distributions.  We estimate that the error made in the total number of binaries by use of this compromise value is of order $5\%$ of the number of binaries.

\section{Conclusions}\label{sec:conc}

In this paper we have presented an advanced Monte Carlo code (MOCCA)
for the evolution of rich star clusters, including most aspects of
dynamical interactions involving binary and single stars, internal
evolution of single and binary stars and the complicated process of
escape in the static tidal field. The direct integration of  few body
encounters  was introduced on the basis of the Fewbody code developed by
\citet{FCZR2004}.  The stellar and binary internal evolution was done
according to the BSE code \citep[]{Hu2000,Hu2002}.  The description of the
escape process was based on the theory presented by \citet{FH2000}.
Thus MOCCA is able to follow all channels of interaction up to
binary-binary encounters, including merging of stars; and  escape is
not immediate any more: stars need time to find their escape route
past  the $L_1$ and $L_2$ Lagrangian points.  The probability of
escape and the probability for interaction are characterized by some
free parameters which were adjusted by comparison of MOCCA and
$N$-body simulation results for systems with large $N$, up to $N=200000$.

It was shown that the free parameters of MOCCA can be
successfully calibrated against $N$-body simulations and that the free
parameters do not depend much on $N$.   MOCCA is not only able to
follow the evolution of the  total mass of the cluster, Lagrangian
radii and the core radius, but also is able to reproduce with
reasonable accuracy distributions of binary parameters and the numbers
of BH-BH binaries and BSS. It also reproduces very well the results
obtained by \citet{Ba2001} for single mass tidally limited systems for
the half-mass time and evolution of potential escapers.  The code is
able to cope with very diverse systems, from single mass, isolated
systems without primordial binaries to multimass, tidally limited
systems with a large fraction of binaries.  It was shown that the
simplified and faster version of MOCCA (without the direct
Fewbody integrator - MOCCA-NoFB) is a method of choice for projects
whose aim is to investigate the evolution of star clusters from the
point of view their global properties.  For other purposes,
particularly when properties of ``peculiar" objects and their
distributions are  of interest, one should use the slower MOCCA
code. It is worth  noting that the MOCCA and MOCCA-NoFB simulations presented in this paper need only about three and two hours, respectively, to be completed on a standard Opteron 2.4Ghz CPU.  

Despite these successes MOCCA still has some known shortcomings, which we summarise here.
\begin{enumerate}
\item{\sl Higher-order multiples}:  It is widely argued that primordial 
triples and higher multiples should be incorporated into simulations along 
with primordial binaries. In any case, hierarchical triples form abundantly 
in binary-binary interactions \citep{mikkola1984}. Such higher-order 
multiples are ignored in the present Monte-Carlo code; they are merely
counted and then artificially disrupted \citep{HyG2012}.  It is
planned to introduce hierarchical triples and higher-order multiples
in a later version of the code. 
\item{\sl Rotation}: the Monte Carlo code is based on spherical symmetry, and 
would require rather fundamental and very difficult reconstruction in order 
to cope with cluster rotation.  Rotation accelerates the rates of core
collapse and mass segregation \citep[e.g.][and references
  therein]{FSK2006,KYLS2008}.  In our models the absence of these rotational effects can be compensated by a modest alteration of the initial conditions.
\item{\sl Static tide}: the effects of tidal shocks have been extensively 
studied  and it would be possible to add these effects as another process altering the energies and angular momenta of the stars in the simulations. The addition of tidal shocks will be more important when modelling Galactic globular clusters than open clusters, which are usually confined inside the Galactic disk. 
\end{enumerate}

Despite these limitations, some of which are difficult to cure, the
MOCCA code
presented in this paper shows its potential power in simulations of real star
clusters, from open clusters to rich globular clusters.  Monte Carlo models are
feasible in a reasonable time (a few days) for the many globular clusters
which are too large for direct $N$-body
simulations, and which will remain so for some time. A Monte
  Carlo code is able to provide data as detailed as an
$N$-body code can.  Only these two methods can provide such comprehensive information.  Even when $N$-body simulations eventually become possible, Monte Carlo models will remain as a quicker way of exploring the parameter space for  large scale $N$-body simulations. 

\section*{Acknowledgements}\label{sec:ack} 
The authors are grateful to John Fregeau for making his Fewbody code
accessible to the public and for very helpful suggestions related to
efficient use of the code. We also thank the referee for his detailed
help in making this paper clearer.  AH was supported by the
Polish National Science Center grant DEC-2011/01/N/ST9/06000. MG was supported by the Polish Ministry of
Sciences and Higher Education through the grant N N203 38036.  He
warmly thanks Douglas Heggie for his hospitality during a visit to
Edinburgh University.  DCH thanks MG for his kind hospitality during a
recent visit to CAMK.

\bsp

\label{lastpage}

\end{document}